\documentclass[12pt]{article}
\usepackage{amssymb}
\usepackage{cite,multirow} 
\usepackage[dvips]{graphicx,epsfig}
\usepackage{graphics}

\def\DJ{{\hbox{D\kern-.8em\raise.15ex\hbox{--}\kern.35em}}}

\begin{document}


\rightline{SISSA 29/2007/EP}
\rightline{LPTENS 07/20}
\vskip 2cm


\centerline{\Huge $BPS$ Partition Functions for Quiver Gauge Theories:}
~\\
\centerline{\Huge Counting Fermionic Operators}
~\\
\renewcommand{\thefootnote}{\fnsymbol{footnote}}
\begin{center}
 \bf Davide Forcella${}^{a,b}$ \footnote{\tt forcella@sissa.it\\
$a$:permanent\\
$b$:temporary} 
\end{center}
~\\
{\small
\begin{center}
${}^a$ International School for Advanced Studies (SISSA/ISAS) \\
and INFN-Sezione di Trieste,\\
via Beirut 2, I-34014, Trieste, Italy\\
~\\
 $^{b}$ Laboratoire de Physique Th\'eorique de l'\'Ecole Normale Sup\'erieure\\
24, rue Lhomond, 75321 Paris Cedex 05, France\\
~\\
$^b$ LPTHE, Universit\'es Paris VI et VII, Jussieu \\
75252 Paris, France
~\\
\end{center}
}

\vskip 0.8in

\begin{abstract}
We discuss a general procedure to obtain $1/2$ $BPS$ partition functions for generic $\mathcal{N}=1$ quiver gauge theories. 
These functions count the gauge invariant operators (bosonic and fermionic), charged under all the global symmetries (mesonic and baryonic), 
in the chiral ring of a given quiver gauge theory. In particular we discuss the inclusion of the spinor degrees of freedom in the 
partition functions.

\end{abstract}

\setcounter{footnote}{0}
\renewcommand{\thefootnote}{\arabic{footnote}}
\vskip -0.8in

\vfill
\newpage

\tableofcontents

\section{Introduction}

In the few past years there were a great effort in the study of quiver gauge theories \cite{gauntlett,kru2,Hanany:2005ve,Hanany:2005ss,Feng:2005gw,Butti:2006nk,Butti:2005vn}. 
Understanding the structure of the spectrum of quiver gauge theories is an important topic in the AdS/CFT correspondence 
and more generically in the study of supersymmetric gauge theories and their moduli spaces. Recently a lot of papers studied 
generating functions counting $\frac{1}{2} BPS$ operators in $\mathcal{N}=1$ quiver $CFT$ \cite{5per5,Forcella:2007wk,Benvenuti:2006qr,Butti:2006au,Martelli:2006yb,Feng:2007ur,Romelsberger:2005eg,Kinney:2005ej,Biswas:2006tj,Mandal:2006tk,Martelli:2006vh,Basu:2006id,Hanany:2006uc,Nakayama:2007jy}. The basic idea is to construct a character associated to the $CFT$ that counts the elements in the chiral ring of 
the theory according to their charges under the various $U(1)$ factors of the global symmetry group. \\
The counting procedure consists in the definition of a set of chemical potentials $\{t_i \}$, $i=1,...,r+a$ associated 
to each $U(1)$ factor in the group $U(1)^{r+a}$ (the abelian torus of the global symmetry group of the theory)\footnote{The quiver gauge theories we are going to consider are obtained as near horizon limit of a system of $N$ $D3$ branes placed at the tip of a conical CY singularity.
 Their gauge group is a product of $SU(N)$ factors and the matter content are chiral bifundamentals fields. 
The global symmetries are divided in two families: the ones coming from the isometries of the compactification manifold, 
and the ones coming from the reduction of the $C_4$ form over topologically non trivial three cycles. The former are the 
``flavor symmetries'' of the field theory and contain an abelian torus $U(1)^r$, ( $r>0$, the conformal symmetry imply the 
presence of the $U(1)$ $R$ symmetry); while the second is the abelian group $U(1)_B^a$ of the baryonic symmetries of the field theory.}, 
and the construction of the character $g(\{t_i\})$, such that once expanded for small values of the $t_i$:
\begin{equation}
g(\{t_i\})= \sum_{i_1,...,i_{r+a}} c_{i_1,...,i_{r+a}}t_1^{i_1}...t_{r+a}^{i_{r+a}}
\end{equation} 
the positive integer coefficients $c_{i_1,...,i_{r+a}}$ tell us how many independent gauge invariant operators there are 
in the chiral ring with that specific set of quantum numbers \footnote{In the paper we will call $g(\{t_i\})$ with four 
different names: partition function, generating function, character, Hilbert series. The first one is the physic literature 
name while the second, third and fourth ones are the mathematical literature names.}.\\
This enumeration problem turned out to be a very powerful toll for the study of the properties of the $CFT$ and the dual 
$AdS_5 \times H$ geometry  \cite{5per5,Forcella:2007wk,Benvenuti:2006qr,Butti:2006au,Martelli:2006yb,Feng:2007ur,Romelsberger:2005eg,Kinney:2005ej,Biswas:2006tj,Mandal:2006tk,Martelli:2006vh,Basu:2006id,Hanany:2006uc,Nakayama:2007jy}. The function $g(\{t_i\})$ 
contains the informations regarding the algebraic equations of the moduli space of the field theory, its dimension, 
the volume of the horizon manifold $H$ and hence the value of the central charge $a$ of the $CFT$, the volumes of all 
the non trivial three cycles inside $H$ and hence of all the $R$ charges of the chiral fields in the field theory and 
the behavior of the $BPS$ spectrum under the non perturbative quantum corrections in the strong coupling regime.\\
It turned out that all the geometric informations are encoded in the scalar part of the chiral ring, and indeed the 
papers in literature focused before on the mesonic scalar chiral ring and after on the more difficult baryonic scalar chiral ring.\\
Right now we have a good understanding of the scalar part of the complete chiral ring (containing all the mesonic and baryonic 
degrees of freedom) for a generic quiver gauge theory \cite{5per5}.\\
In this paper we propose to study the generating functions for the complete chiral ring of generic quiver gauge theories 
containing all the scalar degrees of freedom plus the spinorial degrees of freedom originated by the $W^i_{\alpha}$ Weyl 
spinor superfields associated to the $\mathcal{N}=1$ vector multiplets. \\
The complete generating function for the chiral ring of a given quiver gauge theory gives informations about the complete 
$\frac{1}{2}$ $BPS$ spectrum, and hence the possibility of a statistical studies of the thermodynamical properties of a 
generic strongly coupled $CFT$.\\
We will show how it is possible to obtain the generating functions for the complete chiral ring starting from the generating 
functions for the scalar part. This counting procedure is divided into two parts: the $N=1$ counting, and the generic finite $N$ counting.
As usual it happens that the knowledge of the $N=1$ generating function is enough to implement the finite $N$ counting. 
We will indeed introduce a ``superfield formalism'' that allows to pass from the $N=1$ scalar generating function to the $N=1$ complete one. 
In the chiral ring there are both bosonic and fermionic degrees of freedom. Indeed the bifoundamental fields are scalar superfields and 
hence bosons, while the superfiels $W_{\alpha}^i$ are Weyl spinors and hence fermions. This different statistical behaviour became 
important in the finite $N$ counting. For this reason we will introduce the notion of generalized Plethystic exponential $PE$. 
This is a simple mathematical function that implement the mixed statistic of a system of bosons and fermions. Once we have obtained 
the generating function $g_1( \{ t_i \} )$ for $N=1$, to have the generating function $g_N(\{t_i\})$ counting the chiral ring 
operators for finite values of $N$ we just need to plug $g_1( \{ t_i \} )$ in the generalized Plethystic exponential
\footnote{This is just a schematic expression. The more precise equations will be given in the text of the paper.}:
\begin{equation}
\sum_{N=0}^{\infty} \nu^N g_N (\{ t_i \}) = PE[g_1( \{ t_i \} )]= PE^{\mathcal{B}}[g_1^{\mathcal{B}}( \{ t_i \} )] PE^{\mathcal{F}}[g_1^{\mathcal{F}}( \{ t_i \} )]
\end{equation}
where the $\mathcal{B}$, $\mathcal{F}$ are for bosonic and fermionic statistic respectively; $g_1^{\mathcal{B}}( \{ t_i \} )$ 
is the bosonic part of the $N=1$ generating function, while $g_1^{\mathcal{F}}( \{ t_i \} )$ is the fermionic part of the $N=1$ 
generating function. The function $PE^{\mathcal{B}}[...]$ is the usual  Plethystic exponential used in literature while the $PE^{\mathcal{F}}[...]$ is its fermionic version to be defined in the following.\\
Right now we have powerful tools to compute the $N=1$ scalar generating functions \cite{5per5,Forcella:2007wk,Benvenuti:2006qr,Butti:2006au,Martelli:2006yb,Feng:2007ur,Romelsberger:2005eg,Kinney:2005ej,Biswas:2006tj,Mandal:2006tk,Martelli:2006vh,Basu:2006id,Hanany:2006uc,Nakayama:2007jy} and the finite $N$ complete generating functions is simply obtained using the algorithmic procedure previously explained.\\  
The paper is organized in the following way.\\
In Section \ref{N4} we will start with the simplest example of the $\mathcal{N}=4$ gauge theory. We will construct 
the various generating functions looking directly at the chiral ring structure. This is a simple and explicit example 
to take in mind in the next more abstract Section. In Section \ref{gentheo} we will give a general discussion of the 
structure of the chiral ring of quiver gauge theories. We will introduce the concept of generalized $PE$ and we will 
explain how to compute the generating functions for the complete chiral ring in the case $N=1$ and in the generic finite $N$ case. 
In Section \ref{n4reload}, using the tools developed in the previous Section, we will revisit the computation of the generating 
functions for the $\mathcal{N}=4$ gauge theory and we will make some checks of the proposal. In Section \ref{conifo} 
we will introduce our main example: the gauge theory obtained as near horizon limit of a system of $D3$ brane placed 
at the tip of the conifold singularity. We will construct the $N=1$ and finite $N$ generating functions for the mesonic 
sector of the chiral ring, for the sector with baryonic charge $B=1$, and finally for the complete chiral ring with all 
the charges and all the chiral fields. We will than make some checks of the proposal and in Section \ref{conl} we will conclude.

\section{The $\mathcal{N}=4$ generating functions}\label{N4}

Let us start with the easy explanatory example of the $\mathcal{N}=4$, $U(N)$ gauge theory.
In this theory the basic chiral operators are the three scalars superfields:
\begin{equation}
\phi_i \hbox{  }\hbox{  }\hbox{  } i = 1,2,3
\end{equation}
and the spinor superfiels:
\begin{equation}
W_{\alpha} \hbox{  } \hbox{  }\hbox{  } \alpha = +,-
\end{equation}
From the relation in the chiral ring we have:
\begin{equation}\label{cinrel}
\{W_{\alpha},W_{\beta}\}=0 \hbox{  } \hbox{  }\hbox{  } [W_{\alpha}, \phi _i]=0
\end{equation}
Because the fields $\phi_i$ are bosons while the fields $W_{\alpha}$ are fermions, the first will be represented by 
commuting variables while the latter by anti commuting variables.
From the super potential we have the other relations:
\begin{equation}\label{dynnnrel}
[\phi_i, \phi _j]=0
\end{equation}
We would like to write down a generating function counting all the single trace operators in the chiral ring 
of the $\mathcal{N}=4$ gauge theory.\\
This is easy to obtain if we remember that the generic $1/8$ $BPS$ single trace operators are given by: 
\begin{equation}\label{chi}
\hbox{ Tr }( \phi_1^i\phi_2^j\phi_3^k )\hbox{ , }\hbox{ Tr }( W_{\alpha}\phi_1^i\phi_2^j\phi_3^k )\hbox{ , }\hbox{ Tr }( W_{\alpha} W^{\alpha} \phi_1^i\phi_2^j\phi_3^k )
\end{equation}
Let us introduce the chemical potential $q$ for the dimension of the fields:
\begin{equation}
\phi_i\rightarrow q \hbox{ , } W_{\alpha} \rightarrow q^{3/2} \hbox{ , } W_{\alpha}W^{\alpha}  \rightarrow q^3
\end{equation}
the chemical potential $w$ for the fields $W_{\alpha}$:
\begin{equation}
W_{\alpha} \rightarrow w
\end{equation}
and the chemical potential $\alpha$ for the spin of the chiral fields:
\begin{eqnarray}
W_+ \rightarrow \alpha & , & W_- \rightarrow 1/\alpha 
\end{eqnarray}
Hence the operators in (\ref{chi}) will be counted by the following generating functions\footnote{we will 
put the subscript $1$ to the generating functions because it can be shown that for the mesonic part of 
undeformed quiver gauge theories the generating functions for the multi traces, and hence of the complete 
spectrum of gauge invariant operators, in the case $N=1$ are the same as the generating functions for the single 
traces in the limit $N \rightarrow \infty$ \cite{Benvenuti:2006qr}.}:
\begin{eqnarray}\label{g1bfb}
& & g_1^0(q) = \nonumber\\
& & \sum_{i,j,k} q^{i+j+k} = \sum_{n=0}^{\infty} \frac{(n+2)(n+1)}{2}q^n = \frac{1}{(1-q)^3}\nonumber\\
& & \nonumber\\
& & g_1^w(q,w,\alpha) = \nonumber\\
& & \sum_{i,j,k} q^{3/2+i+j+k}w \Big(\alpha + \frac{1}{\alpha}\Big) = \sum_{n=0}^{\infty} \frac{(n+2)(n+1)}{2}q^{3/2+n}w\Big(\alpha + \frac{1}{\alpha}\Big)= \frac{q^{3/2} w \big(\alpha + \frac{1}{\alpha}\big)}{(1-q)^3}\nonumber\\
& & \nonumber\\
& & g_1^{w^2}(q,w) = \nonumber\\
& & \sum_{i,j,k} q^{3+i+j+k}w^2 = \sum_{n=0}^{\infty} \frac{(n+2)(n+1)}{2}q^{3+n}w^2 = \frac{ q^3w^2  }{(1-q)^3} 
\end{eqnarray}
Hence the complete generating function counting all the single trace operators (\ref{chi}) is:
\begin{equation}
g_1(q,w,\alpha)=g_1^0(q)+ g_1^w(q,w,\alpha)+ g_1^{w^2}(q,w)= \frac{1 + q^{3/2} w \big(\alpha + \frac{1}{\alpha}\big)+ w^2 q^3}{(1-q)^3}
\end{equation}
Right now the counting was done in the limit $N \rightarrow \infty$ and in the single trace sector i.e. 
without taking into account the relations among gauge invariants coming from the fact that for finite $N$ 
one can rewrite some of the operators in terms other operators with lower dimensions.\\
To understand the finite $N$ counting we can look to equations (\ref{cinrel},\ref{dynnnrel}). These ones tell 
that one can diagonalize at the same time all the $\phi_i$ and all the $W_{\alpha}$. Hence we are left with a 
system of $3N$ bosonic eigenvalues and $2N$ fermionic eigenvalues. 
Using the Bose-Einstein and Fermi-Dirac statistic we can write down the function $g(q,w,\alpha;\nu)$ that 
generate the partition functions $g_N(q,w,\alpha)$ with fixed $N$ counting all the $\frac{1}{8}BPS$ gauge 
invariant operators (single and multi traces)\cite{Kinney:2005ej}:
\begin{eqnarray}\label{bosferN}
& & g(q,w,\alpha;\nu)=\sum_{N=0}^{\infty} \nu^N g_N(q,w,\alpha)=\nonumber\\
& & \prod_{n=0}^{\infty}\frac{(1+\nu \hbox{ }w\hbox{ } \alpha\hbox{ } q^{3/2+n})^{ \frac{(n+2)(n+1)}{2}}(1+ \nu \hbox{ }w\hbox{ } \frac{1}{\alpha}\hbox{ } q^{3/2+n})^{ \frac{(n+2)(n+1)}{2}}}{(1-\nu\hbox{ } q^n)^{\frac{(n+2)(n+1)}{2}}(1- \nu \hbox{ }w^2\hbox{ } q^{3+n})^{\frac{(n+2)(n+1)}{2}}} \nonumber\\
\end{eqnarray}
In the next Section we will discuss how to construct in general the generating functions for the chiral ring 
of a given quiver gauge theory. After this we will revisit the $\mathcal{N}=4$ case and we will make some checks 
of the prescription.

\section{The general approach to the complete generating functions for quiver gauge theories}\label{gentheo}

Let us try to motivate our approach to the complete partition function for the chiral ring of a given gauge theory. 
The gauge theories we are going to consider are obtained as near horizon limit of a system of $N$ $D3$ branes at the 
tip of a $CY$ conical singularity. These theories have a set of flavor symmetries (we include in this class also the 
always present $R$ symmetry) $U(1)^r$ and a set of baryonic symmetries $U(1)_B^a$. We would like to write down a 
generating function counting all the operators in the chiral ring according to their charges under the global symmetries 
of the theory. From the computation of the generating functions in the scalar chiral ring \cite{Forcella:2007wk, Benvenuti:2006qr,Butti:2006au,Feng:2007ur,Hanany:2006uc,5per5} we know that the knowledge of the generating function for a single $D3$ brane is enough 
to compute the generating functions for an arbitrary number $N$ of $D3$ branes\footnote{this is due to the fact that the 
gauge invariant operators for fixed $N$ are nothing else that the $N$ times symmetric product of the ones for $N=1$}. 
The last ones are indeed obtained by the former using some combinatorial tools. For this reason one has a well defined 
notion of single particle Hilbert space ($N=1$) and multi particle Hilbert spaces ($N > 1$ ), that we will use in the following. \\
A generic quiver gauge theory has the following set of chiral fields:
\begin{equation}\label{scachi}
X_{i,j}^{e_{ij}}
\end{equation}
these ones are scalar bifundamental superfields transforming in the fundamental of the $SU(N)_i$ gauge group and in 
the anti fundamental of the $SU(N)_j$ gauge group. The $e_{ij}=1,...,E_{ij}$ label the number of fields between the 
$i$-th gauge group and the $j$-th gauge group;
\begin{equation}\label{spichi}
W_{\alpha}^i
\end{equation}
these ones are the Weyl spinor superfields associated to the vector supermultiplet of the $SU(N)_i$ gauge group, 
$\alpha = +,-$ label the spin state and the label $i$ goes over all the $G$ gauge groups $i=1,...,G$.\\
In the chiral ring they must satisfy the ``symmetry'' relations\footnote{see for example \cite{Casero:2003gf}}:
\begin{eqnarray}\label{symrel}
& & \{ W_{\alpha}^i, W_{\beta}^i \}=0 \nonumber\\
& & W_{\alpha} ^i X_{i,j}^{e_{ij}} = X_{i,j}^{e_{ij}} W_{\alpha}^j 
\end{eqnarray}
and the dynamical relations:
\begin{equation}\label{dynrel}
\frac{\partial \mathcal{W}}{\partial  X_{i,j}^{e_{ij}}}= 0 
\end{equation}
Where $\mathcal{W}$ is the superpotential of the theory under consideration.\\
The gauge invariant operators constructed with scalar fields $X_{i,j}^{e_{ij}}$ define a complex ambient space. 
The dynamical equations (\ref{dynrel}) define an algebraic variety in this space and hence they give constraints 
inside the chiral ring that are usually hard to deal with. Thanks to the recent works \cite{Forcella:2007wk, Benvenuti:2006qr,Butti:2006au,Feng:2007ur,Hanany:2006uc,5per5} the problem of counting the scalar part of the chiral ring, and hence how to deal 
with (\ref{dynrel}) is right now under control. What we want to show here is how, starting from the knowledge 
of the scalar partition function, one can write down the complete $\frac{1}{2}$ $BPS$ partition function for a generic quiver gauge theory.\\
The generic gauge invariant operator inside the chiral ring will be constructed in the following way:
given a pair of gauge groups $(x,z),\ x,z=1,...,G$, we call a gauge invariant of type $(x,z)$ a gauge invariant of the form
\begin{equation}\label{ginv}
\epsilon_{x}^{k_1,...,k_N} ({\bf O}_{I_1}^{(x,z)})_{k_1}^{l_1} .... ({\bf O}_{I_N}^{(x,z)})_{k_N}^{l_N}\epsilon^{z}_{l_1,...,l_N}
\end{equation}
where $({\bf O}_I^{(x,z)})_{k}^{l}$ denotes a string of elementary chiral fields $X_{i,j}^{e_{ij}}$, $W_{\alpha}^i$ with all gauge indices
contracted except two indices, $k$ and $l$, corresponding to the gauge groups $(x,z)$. The index $I$ runs over all possible strings of 
elementary fields
with these properties.
The full set of gauge invariant operators is obtained by arbitrary products of the operators in (\ref{ginv}). Using the tensor
relation
$$\epsilon^{k_1,...,k_N}\epsilon_{l_1,...,l_N} = \delta^{k_1}_{[l_1}\cdots \delta^{k_N}_{l_N]} $$
some of these products of determinants are equivalent and some of these are actually equivalent
to mesonic operators made only with traces.\\
From the first equation in (\ref{symrel}) it is easy to understand that the fields $W_{\alpha}^i$ are the fermionic degrees of freedom 
of the theory, and they can appear in the chiral operator $({\bf O}_I^{(x,z)})_{k}^{l}$ alone or at most in couple in the antisymmetric 
combination: $W_{\alpha}^i W^{\alpha}_i = W_{+}^i W_{-}^i - W_{-}^i W_{+}^i$. The second equation in (\ref{symrel}) 
tells that the position of the $W_{\alpha}^i$ spinor fields inside the chiral string $({\bf O}_I^{(x,z)})_{k}^{l}$ does not matter. 
From now on we will call the ``single particle Hilbert space'' the space spanned by the operators $({\bf O}_I^{(x,z)})_{k}^{l}$. 
This space is the total space of the gauge invariant operators in the case $N=1$.\\
Once we know the spectrum of the operators inside the scalar chiral ring of the single particle Hilbert space 
(i.e. all the $({\bf O}_I^{(x,z)})_{k}^{l}$ that does not contain the $W^i_{\alpha}$ field), the generic operators come in class: 
there are the ones of the scalar chiral ring, the same ones with the insertion of one field $W_{\alpha}$, 
and the same ones of the scalar chiral ring with the insertion of the antisymmetric combination  $W_{\alpha}^i W^{\alpha}_i$. 
Thanks to the ``commutativity'' properties of the fields  $X_{i,j}^{e_{ij}}$ and $W_{\alpha}$ (\ref{symrel}), 
we do not have to specify the gauge groups $i$ the $W_{\alpha}$ fields belong to. It is enough to 
pick up a representative $W_{\alpha}$ for all the gauge groups. The result of this operation is that 
we would be able to easily write down the complete partition function for the $\frac{1}{2}$ $BPS$ 
states of a given $\mathcal{N}=1$ quiver gauge theory with gauge group given by a product of $SU(N)$ factors, 
times a diagonal overall $U(1)$ factor. The latter comes from the operators that factorize in products of the type Tr$(W_{\alpha}) (...)$. \\
To every operators $({\bf O}_I^{(x,z)})_{k}^{l}$ we can associate a state in the single particle Hilbert space.
The single particle space of the complete chiral ring would be spanned by the states:
\begin{equation}\label{spauno}
|m_1,...,m_r,B_1,...B_a,\mathcal{S}>
\end{equation}
where $m_i$ are the charges under the $T^r$ abelian torus inside the generically non abelian 
global flavor group of the gauge theory, $B_j$ are the charges under the $U(1)^a_B$ baryonic 
symmetry of the theory, and $\mathcal{S}=\mathcal{B},\mathcal{F}$ label the statistic of the state.\\
Now that we have understood the structure of the chiral ring we can divide the generic one particle state (\ref{spauno}) 
in three classes according to the number of $ W_{\alpha}^i$ fields.
\begin{eqnarray}\label{symrelo}
& & |m_1,...,m_r,B_1,...B_a,0> \hbox{  } \in \hbox{  } \mathcal{H}^{\mathcal{B}}_{0,1} \nonumber\\
& & |m_1,...,m_r,B_1,...B_a,w^i> \hbox{  } \in \hbox{  } \mathcal{H}^{\mathcal{F}}_{w,1} \nonumber\\
& & |m_1,...,m_r,B_1,...B_a,w^iw^i> \hbox{  } \in \hbox{  } \mathcal{H}^{\mathcal{B}}_{w^2,1}
\end{eqnarray}
where the first and the third state are bosonic states ($\mathcal{B}$), while the second one is fermionic ($\mathcal{F}$), 
the subscripts $0,w,w^2$ label the presence of $W_{\alpha}$ fields, and the second subscript label the number of particle $n$, 
in this case ( single particle ) $n=1$.\\
Let us now pass to the generic $N$ case and hence to the multi particle Hilbert space. 
Thanks to the division (\ref{symrelo}) the Fock space of the chiral ring ($c$.$r$.) will be divided into three parts:
\begin{equation}\label{fockchiral}
\mathcal{F}^{c.r.}= \mathcal{F}^{\mathcal{B}}_0 \otimes  \mathcal{F}^{\mathcal{F}}_w  \otimes \mathcal{F}^{\mathcal{B}}_{w^2}
\end{equation}
where the three factors on the right hand side are the Fock spaces associated to the one particle Hilbert spaces definite in (\ref{symrelo}):
\begin{equation}\label{fock3}
\mathcal{F}^{\mathcal{B}}_0=  \bigoplus_n^{\infty}\mathcal{H}^{\mathcal{B}}_{0,n} \hbox{  } \hbox{ , } \hbox{  } \mathcal{F}^{\mathcal{F}}_w=  \bigoplus_n^{\infty}\mathcal{H}^{\mathcal{F}}_{w,n}  \hbox{  } \hbox{ , } \hbox{  } \mathcal{F}^{\mathcal{B}}_{w^2}= \bigoplus_n^{\infty}\mathcal{H}^{\mathcal{B}}_{w^2,n} 
\end{equation}
where the $\mathcal{B}$ means the Fock space of a symmetrized tensor product of the single particle states, 
while the $\mathcal{F}$ means the Fock space of an anti symmetrized tensor product of the single particle states. 
The chiral ring Fock space decomposes in the following way:
\begin{equation}\label{fockchiraldec}
\mathcal{F}^{c.r.}= \bigoplus_N^{\infty} \bigoplus_{p+q+r=N}^{\infty} \mathcal{H}^{\mathcal{B}}_{0,p} \otimes  \mathcal{H}^{\mathcal{F}}_{w,q} \otimes \mathcal{H}^{\mathcal{B}}_{w^2,r}=\bigoplus_N^{\infty} \mathcal{H}^{c.r.}_N
\end{equation}
This means that we have to deal with a multi particle space that is composed by bosons and fermions. 
If we want to count the operators at finite $N$ we must introduce mathematical functions that implement 
the bosonic and the fermionic statistic.\\
Let us define the one particle generating function:
\begin{equation}
g_1(q)=\sum_{n=0}^{\infty} a_n q^n
\end{equation}
counting $\frac{1}{2}BPS$ operators according for example to their dimension: the integer numbers $a_n$ 
tells us how many operators we have with dimension $n$.\\
For the finite $N$ counting we have to implement the right statistic.\\
For the bosonic part of the spectrum we have the usual Plethystic function \cite{Forcella:2007wk, Benvenuti:2006qr,Butti:2006au,Feng:2007ur,Hanany:2006uc,5per5}, that from now on we will call it the bosonic Plethystic function ( $PE^{\mathcal{B}}$ ): 
\begin{equation}\label{PEBB}
\prod_{n=0}^{\infty}\frac{1}{(1- \nu  q^n)^{a_n}}= PE^{\mathcal{B}}[g_1(q)]\equiv \exp\Big( \sum_{k=1}^{\infty} \frac{\nu^k}{k}g_1(q^k)\Big)=\sum_{N=0}^{\infty}\nu ^N g_N(q)
\end{equation}
This function takes a certain generating function $g_1(q)$ and generates new partition functions $g_N(q)$ 
counting all the possible $N$ times symmetric products of the constituents of $g_1(q)$, implementing in 
this way the bosonic statistic, as it is possible to see from the left hand side of (\ref{PEBB}).\\
For the fermionic part it is useful to introduce the fermionic Plethystic function ( $PE^{\mathcal{F}}$ )\cite{Feng:2007ur}:
\begin{equation}\label{PEFF}
\prod_{n=0}^{\infty}(1+ \nu  q^n)^{a_n}=PE^{\mathcal{F}}[g_1(q)]\equiv \exp\Big( \sum_{k=1}^{\infty}(-)^{k+1} \frac{\nu^k}{k}g_1(q^k)\Big)=\sum_{N=0}^{\infty}\nu ^N g_N(q)
\end{equation}
This function generate the partition functions $g_N(q)$ counting all the possible $N$-times anti symmetric 
products of the objects counted by $g_1(q)$, and implements in this way the fermionic statistic, as it is 
possible to see from the left hand side of (\ref{PEFF}).\\
Once we have defined the three basic generating functions (one for each one of the states in (\ref{symrelo})), 
the counting problem for a generic quiver gauge theories translates in counting the states in the Fock space 
defined in (\ref{fockchiraldec}).\\
Let us define the following chemical potentials:
\begin{itemize}
\item{$q=(q_1,...,q_r)$ labels the flavor charges;}
\item{$b=(b_1,...,b_a)$ labels the baryonic charges;}
\item{$\alpha$ labels the spin}
\item{$w$ labels the number of $W_{\alpha}$ fields.}
\end{itemize}
For each of the sectors in (\ref{symrelo}) we can associate a generating function, counting the single 
particle operators with a fixed set of baryonic charges $B=(B_1,...,B_a)$:
\begin{equation}
g_{1,B}^0(q) \hbox{  } \hbox{ , } \hbox{   } g_{1,B}^w(q,w,\alpha) \hbox{  } \hbox{ , } \hbox{  } g_{1,B}^{w^2}(q,w)
\end{equation}
With these definitions the finite $N$ counting is implemented by the following total generating function of the chiral ring:
\begin{eqnarray}\label{peppa}
& &  g^{c.r.}(q,w,\alpha,b;\nu)= \sum_{N=0}^{\infty} \nu^N g_N^{c.r.}(q,w,\alpha,b)= \nonumber\\
& &\sum_{B}b^B m(B)PE^{\mathcal{B}}[g^{0}_{1,B}(q)]PE^{\mathcal{F}}[g^{w}_{1,B}(q,w,\alpha) ]PE^{\mathcal{B}}[g^{w^2}_{1,B}(q,w,\alpha) ]\nonumber\\\end{eqnarray}
where the expression $b^B$ means $b_1^{B_1}...b_a^{B_a}$, and $m(B)$ is what we call the multiplicities: 
the number of equal generating functions with the same set of baryonic charges and distinguished just by 
the field theory content \cite{5per5}. The meaning of (\ref{peppa}) is: with the $PE$ functions we implement 
the right statistic for the various states and then we sum over all the possible sectors with fixed baryonic 
charges taking into account the possible presence of multiplicities.\\
To obtain the generating function for the chiral ring with fixed number of branes $N$ one have just to take the $N$ 
times derivatives of (\ref{peppa}) with respect to the parameter $\nu$:
\begin{equation}
g_N^{c.r.}(q,w,\alpha,b) = \frac{1}{N !}\frac{\partial^N g^{c.r}(q,w,\alpha,b;\nu)}{\partial \nu^N} \Big|_{\nu = 0 }
\end{equation}
What we miss is to construct the generating functions for the two sectors of the Hilbert space containing the 
fields $W_{\alpha}$ once we know $g_{1,B}^0(q)$. This one is an easy task and can be solved in an elegant way 
introducing a superfield formalism. Let us introduce the usual set of anti commuting variables $\theta_{\alpha}$ such that:
\begin{equation}
\{ \theta _{\alpha} , \theta _{\beta} \} = 0 
\end{equation}
the dimension of the theta variables is $-3/2$ and we will label it with $q^{-3/2}$, in addition they carry a 
spin degrees of freedom that we will label with $\alpha$ for the $1/2$ spin case and $1/ \alpha$ for $-1/2$ spin case. 
As explained in \cite{Butti:2006au} the generic chiral gauge invariant operators constructed with only the scalar 
superfields of the theory is an $N$-times symmetric product of $N$ chiral fields ``building blocks'': the $({\bf O}_I^{(x,z)})_{k}^{l}$ 
fields without insertion of $W^i_{\alpha}$. For simplicity from now on we will call $\phi_B^m$ the generic scalar building block, 
where with $m$ we mean the specific set of $r$ flavor charges and with $B$ the specific set of $a$ baryonic charges of the operator. 
Given the relation inside the chiral ring and the decomposition of the Hilbert space of the chiral ring we propose a superfield 
formalism in which the super chiral fields are generically given by:   
\begin{equation}\label{superfc}
\Phi^m_B = \phi_B^m + \theta_{\alpha} (W^{\alpha}  \phi_B^m)  +  \theta_{\alpha} \theta^{\alpha}(W_{\alpha} W^{\alpha} \phi_B^m )
\end{equation}
Introducing the chemical potential $w$ counting the number of $W_{\alpha}$ fields, it is by now clear that the complete 
generating function for $N=1$ is 
\begin{equation}\label{par1B}
g_{1,B}(q,w,\alpha)=g_{1,B}^0(q)\Big( 1+q^{3/2}\hbox{ }w\hbox{ }\Big( \alpha + \frac{1}{\alpha} \Big) + q^3 \hbox{ } w^2 \hbox{ }\Big)
\end{equation} 
Once we have defined (\ref{par1B}) we can safely divide it in the fermionic part and the bosonic one:
\begin{equation}\label{div}
g_{1,B}(q,w,\alpha)=g_{1,B}^0(q) +  g_{1,B}^w(q,w,\alpha) + g_{1,B}^{w^2}(q,w^2) 
\end{equation}
The statistical behavior of the field (\ref{superfc}) will be a mixed bosonic and fermionic statistic and this 
fact is implemented by the generalized $PE$ function:
\begin{equation}\label{loca}
PE[g_{1,B}(q,w,\alpha)]=PE^{\mathcal{B}}[g_{1,B}^0(q)]PE^{\mathcal{F}}[g_{1,B}^w(q,w,\alpha)]PE^{\mathcal{B}}[g_{1,B}^{w^2}(q,w^2)]
\end{equation}
Let us now take the first derivatives of this expression and look at the form of the finite $N$ generating functions:
\begin{equation}
g_{1,B}(q,w,\alpha) = \partial _{\nu} PE[g_{1,B}(q,w,\alpha)] \Big|_{\nu=0}= g_{1,B}^0(q) + g_{1,B}^w(q,w,\alpha) + g_{1,B}^{w^2}(q,w)\end{equation}
This expression clearly reproduce the $N=1$ counting we started with.\\
The two times derivatives is more interesting: here we can start observing the mixed statistic. 
\begin{eqnarray}\label{PEexpN2}
& & g_{2,B}(q,w,\alpha) = \frac{1}{2} \partial _{\nu}^2 PE[g_{1,B}(q,w,\alpha)] \Big|_{\nu=0} = \nonumber\\
& & \frac{1}{2} \Big( \big( g_{1,B}^0(q^2) + [g_{1,B}^0(q)]^2 \big) + \big( g_{1,B}^{w^2}(q^2,w^2) + [g_{1,B}^{w^2}(q,w)]^2 \big) + 2 g_{1,B}^0(q) g_{1,B}^{w^2}(q,w) + \nonumber\\
& & \big( - g_{1,B}^w(q^2,w^2,\alpha^2) + [g_{1,B}^w(q,w,\alpha)]^2 \big) + 2 \big( g_{1,B}^0(q) + g_{1,B}^{w^2}(q,w) \big) g_{1,B}^w(q,w,\alpha)  \Big) \nonumber\\
\end{eqnarray}
Let us comment factor by factor the equation (\ref{PEexpN2}):
the first two factors take into account the bosonic statistic of the scalar part of the chiral ring, 
and in the usual partition functions written in literature that all. Here instead we have many more terms. 
The second two factors are due to the bosonic statistic of the part of the chiral ring with the insertions 
of the field $W_{\alpha}W^{\alpha}$. The third factor describe the mixing between the two bosonic sectors of the chiral ring. 
The fourth two factors is probably the most interesting one: it has the same form of the first and the second, 
but it has one minus sign more: this is due to the fermionic behavior of $g_{1,B}^w(q,w,\alpha)$. 
This factor implement the fermionic statistic of the operators in the chiral ring with the insertions 
of the field $W_{\alpha}$. The last two factors are due to the mixing between the bosonic and fermionic part of the chiral ring.\\
Before passing to some checks and examples of the general proposal, we want to make a 
comment regarding the $BPS$ mesonic branch of the chiral ring for generic quiver gauge theories. 
Thanks to the techniques developed in \cite{Benvenuti:2006qr} we can easily find the generating 
functions with $N=1$ for the scalar mesonic chiral ring of a quiver gauge theory counting the 
gauge invariant operators according to their dimensions:
\begin{equation}\label{g100}
g_{1,0}^0(q)=\sum_{n=0}^{\infty} a_n q^n
\end{equation}
Once we know the $a_n$ factors in (\ref{g100}), the general form for the function $g_0(q,t,\alpha;\nu)$ 
generating the Hilbert series for the finite $N$ mesonic counting is:
\begin{equation}
g_0(q,w,\alpha;\nu)=\sum_{\nu = 0}^{\infty} \nu ^N g_{N,0}(q,w,\alpha)=\prod_{n=0}^{\infty}\frac{(1+\nu \hbox{  }w\hbox{  } \alpha\hbox{  }
 q^{3/2+n})^{a_n}(1+ \nu \hbox{  }w \hbox{  }\frac{1}{\alpha}\hbox{  } q^{3/2+n})^{a_n}}{(1-\nu \hbox{  } q^n)^{a_n}(1- \nu \hbox{  } w^2 \hbox{  }q^{3+n})^{a_n}} \nonumber\\
\end{equation}

\section{The $\mathcal{N}=4$ case revisited}\label{n4reload}

In this Section we will very briefly review the case $\mathcal{N}=4$ using the technology developed in the previous Section.\\
The basic idea is to pass from the $\delta_i^j$ invariant tensor, that does not depend on the number of colors $N$, 
to the $\epsilon_{i_1,...,i_N}$ invariant tensor that has built in the dependence on $N$. In this way the basic 
generating function counting the gauge invariant operators in the chiral ring for $N=1$ is the one counting the 
single trace in the limit $N \rightarrow \infty $ and all the counting for finite $N$ can be obtained starting 
from the single trace $N \rightarrow \infty $ generating function.\\Following the prescription of the previous 
Section we have just to compute the generating function $g_1^0(q)$ for $N=1$. Using the equivariant index theorem 
this is easy to compute in general, and in the particular case of $\mathcal{N}=4$ this function counts the holomorphic 
functions on $\mathbb{C}^3$ according to their degree and it is exactly the first one in (\ref{g1bfb}):
\begin{equation}
g_1^0(q) = \sum_{n=0}^{\infty}\frac{(n+1)(n+2)}{2}q^n= \frac{1}{(1-q)^3}
\end{equation}
Now we must implement the superfield prescription and write:
\begin{equation}
g_1(q,w,\alpha)=g_1^0(q)\Big( 1+q^{3/2} w \Big( \alpha + \frac{1}{\alpha} \Big) + q^3 w^2\Big)
\end{equation}
which is exactly the one obtained in Section \ref{N4}.\\
To implement the finite $N$ counting we have just to use the generalized $PE$:
\begin{equation}\label{genN4}
g(q,w,\alpha;\nu)=PE[g_1(q,w,\alpha)]=PE^{\mathcal{B}}[g_1^0(q) + g_1^{w^2}(q,w)]PE^{\mathcal{F}}[g_1^{w}(q,w,\alpha)]
\end{equation}
One can easily checks that (\ref{genN4}) reproduce the equation (\ref{bosferN}). \\
Now we want to take the first few derivatives of (\ref{genN4}) and looking for the corresponding gauge invariant operators. 

\subsection{Comparison with the field theory}

Let us start with $N=1$ and expand the corresponding generating function organizing the various terms in powers of $q$:
\begin{eqnarray}\label{expn1n4}
g_1(q,w,\alpha) = 1 + 3 q + \Big( \frac{w}{\alpha} + w \alpha \Big) q^{3/2} + 6 q^2 + \Big( 3\frac{w}{\alpha} + 3 w \alpha \Big) q^{5/2} + (10 + w^2) q^3 + ...\nonumber\\
\end{eqnarray}
The operators in field theory corresponding to the terms in (\ref{expn1n4}) are \footnote{Although we are counting 
the operators in the case N=1, we decided to look at the single trace operators in the limit $N\rightarrow \infty$, 
because these are the same operators we considered in Section \ref{N4}, and as previously explained the two counting 
procedures are equal.}
:
\begin{eqnarray}
1 &\rightarrow& \mathbb{I} \hbox{ : } \hbox{1 operator}\nonumber\\
3 q &\rightarrow& \hbox{ Tr } (\phi _i )  \hbox{ : } \hbox{3 operators}\nonumber\\
\Big( \frac{w}{\alpha} + w \alpha \Big) q^{3/2} &\rightarrow& \hbox{ Tr } (W _{\alpha}) \hbox{ : } \hbox{1 Weyl spinor operator}\nonumber\\
 6 q^2  &\rightarrow&  \hbox{ Tr } (\phi _i \phi_j) \hbox{ : } \hbox{6 operators}\nonumber\\ 
\Big( 3\frac{w}{\alpha} + 3 w \alpha \Big) q^{5/2} &\rightarrow&  \hbox{ Tr } (W _{\alpha} \phi_i) \hbox{ : } \hbox{3 Weyl spinor operators}\nonumber\\10 q^3 &\rightarrow& \hbox{ Tr } (\phi _i \phi_j \phi_k)  \hbox{ : } \hbox{10 operators}\nonumber\\ 
w^2 q^3  &\rightarrow& \hbox{ Tr } (W _{\alpha} W^{\alpha}) \hbox{ : } \hbox{1 operator}
\end{eqnarray}
Let now pass to the more interesting case of $N=2$. \\
\begin{eqnarray}\label{g2c3nnex}
& & g_2(q,w,\alpha)=\nonumber\\
& & \frac{(q^{3/2} w + \alpha )(1 + q^{3/2}w \alpha )(q^{5/2}(3 + q^2)w + (1 + 3 q^2)(1 + q^3 w^2) \alpha + q^{5/2}(3 + q^2)w \alpha^2)}{(1 - q)^6(1 + q)^3 \alpha^2}\nonumber\\
\end{eqnarray}
The expansion of (\ref{g2c3nnex}) organizes in powers of $q$:
\begin{eqnarray}\label{g2c3}
g_2(q,w,\alpha) &=& 1 + 3 q + \Big( \frac{w}{\alpha} + w \alpha \Big) q^{3/2} + 12 q^2 + \Big( 6\frac{w}{\alpha} + 6 w \alpha \Big) q^{5/2} + (28 + 2 w^2) q^3 + \nonumber\\ 
 & & \Big( 21 \frac{w}{\alpha} + 21 w \alpha \Big) q^{7/2} +  \Big( 66 + 12 w^2 + 3 \frac{w^2}{\alpha^2} + 3 w^2 \alpha^2 \Big) q^4 + ...
\end{eqnarray}
The first three terms in (\ref{g2c3}) are the same as the ones for $N=1$, starting from dimension $2$ 
we have new contributions to the generating function:
\begin{eqnarray}
 12 q^2  &\rightarrow& \hbox{ Tr } (\phi _i \phi_j) \hbox{ , } \hbox{ Tr } (\phi _i)\hbox{ Tr }( \phi_j): \nonumber\\
& & \hbox{ $6 + 6$ operators}\nonumber\\
\Big( 6\frac{w}{\alpha} + 6 w \alpha \Big) q^{5/2}  &\rightarrow& \hbox{ Tr } (W _{\alpha} \phi_i) \hbox{ , } \hbox{ Tr } (W _{\alpha}) \hbox{ Tr }(\phi_i): \nonumber\\
& & \hbox{$6+6$ Weyl spinor operators} \nonumber\\
\nonumber\\
28 q^3 &\rightarrow& \hbox{ Tr } (\phi _i \phi_j \phi_k) \hbox{ , } \hbox{ Tr } (\phi _i \phi_j) \hbox{ Tr } (\phi_k): \nonumber\\
&  & \hbox{$10 +18$ operators}\nonumber\\
2 w^2 q^3 &\rightarrow& \hbox{ Tr } (W _{\alpha} W^{\alpha}) \hbox{ , } \hbox{ Tr } (W _{+})\hbox{ Tr }( W_{-}):\nonumber\\ 
& & \hbox{$1 +1$ operators}\nonumber\\  
 \Big( 21 \frac{w}{\alpha} + 21 w \alpha \Big) q^{7/2} &\rightarrow& \hbox{ Tr } (W _{\alpha} \phi_i \phi_j) \hbox{ , } \hbox{ Tr } (W _{\alpha}) \hbox{ Tr }(\phi_i \phi_j) \hbox{ , } \hbox{ Tr } (W _{\alpha} \phi_i) \hbox{ Tr } (\phi_j): \nonumber\\ 
& & \hbox{$6 + 6 + 9 $ Weyl spinor operators}\nonumber\\ 
 66 q^4  &\rightarrow& \hbox{ Tr } (\phi _i \phi_j \phi_k \phi_l) \hbox{ , } \hbox{ Tr } (\phi _i \phi_j)\hbox{ Tr }( \phi_k \phi_l) \hbox{ , } \hbox{ Tr } (\phi _i) \hbox{ Tr } (\phi_j \phi_k \phi_l):\nonumber\\ 
& & \hbox{ $15 + 21 + 30$ operators}\nonumber\\
 12 w^2 q^4  &\rightarrow& \hbox{ Tr } (W _{\alpha} W^{\alpha} \phi_i) \hbox{ , } \hbox{ Tr } (W _{\alpha} W^{\alpha}) \hbox{ Tr }( \phi_i) \hbox{ , } \hbox{ Tr } (W _{+}\phi_i) \hbox{ Tr }(W_{-}) \hbox{ , } \nonumber\\
& & \hbox{ Tr } (W _{-}\phi_i) \hbox{ Tr }(W_{+}): \nonumber\\
& & \hbox{$3 + 3 + 3 + 3$ operators}\nonumber\\  
\Big(3 \frac{w^2}{\alpha^2} + 3 w^2 \alpha^2 \Big) q^4 &\rightarrow& \hbox{ Tr } (W _{+}\phi_i) \hbox{ Tr }(W_{+}) \hbox{ , } \hbox{ Tr } (W _{-}\phi_i) \hbox{ Tr }(W_{-}): \nonumber\\
& & \hbox{ $3+3$ operators}
\end{eqnarray}
It is easy to see that the operators are counted in the right way.

\section{The conifold}\label{conifo}

At this point it is useful to study in detail a less trivial example containing almost all the properties 
of the generic case: let us discuss the conifold gauge theory chiral ring.\\
The gauge theory has the gauge group $SU(N)_1 \times SU(N)_2$.
The basic chiral fields are the four scalar superfields: 
\begin{equation}
A_i \hbox{ , } B_i \hbox{  }\hbox{  } i = 1,2
\end{equation}
The $A_i$ fields transform in the fundamental of $SU(N)_1$ and the anti fundamental of $SU(N)_2$, 
while the $B_i$ fields transform in the fundamental of $SU(N)_2$ and the anti fundamental of $SU(N)_1$; 
and the chiral spinor superfields for the two factors of the gauge group $SU(N)_1 \times SU(N)_2$:
\begin{equation}
W_{\alpha}^1 \hbox{ , } W_{\alpha}^2 \hbox{  }\hbox{  } \alpha = +,-
\end{equation}
The cinematical relations in the chiral ring are:
\begin{equation}\label{relcon}
\{W_{\alpha}^i,W_{\beta}^i\}=0 \hbox{ , } W_{\alpha}^1 A_i = A_i W_{\alpha}^2 \hbox{ , } W_{\alpha}^2 B_i = B_i W_{\alpha}^1
\end{equation}
while the dynamical ones coming from the super potential are:
\begin{equation}
B_1 A_i B_2= B_2 A_i B_1 \hbox{ , } A_1 B_i A_2 = A_2 B_i A_1
\end{equation}

\subsection{The mesonic chiral ring}

Let's start from the easy case of the mesonic chiral ring.\\
The generating function for the scalar part of the chiral ring is\footnote{to be in line with the 
literature in the generating functions for the conifold we will use the chemical potential $q$ to 
label the $R$ charge of the chiral fields: $R_{A_i}=R_{B_i}=1/2$, $R_{W_{\alpha}}=1$.}\cite{Benvenuti:2006qr}:
\begin{equation}
g_{1,0}^0(q) = \sum_{n=0}^{\infty} (1+n)^2 q^n =\frac{(1+q)}{(1-q)^3}
\end{equation}
Using the prescription of Section \ref{gentheo} the complete $N=1$ mesonic generating function is:
\begin{equation}\label{n1conc}
g_{1,0}(q,w,\alpha)= g_{1,0}^0(q)\Big( 1+q w \Big( \alpha + \frac{1}{\alpha} \Big) + q^2 w^2 \Big)
\end{equation}
Now we have just to apply the $PE$ formalism and we obtain the mesonic generating function for finite $N$:
\begin{eqnarray}
& & g_0(q,w,\alpha;\nu)= PE^{\mathcal{B}}[g_{1,0}^0(q)+g_{1,0}^{w^2}(q,w)]PE^{\mathcal{F}}[g_{1,0}^w(q,w,\alpha)]=\nonumber\\
& & \prod_{n=0}^{\infty}\frac{(1 + \nu \hbox{ }w\hbox{ } \alpha\hbox{ } q^{1+n})^{(n+1)^2}(1 + \nu\hbox{ } w\hbox{ } \frac{1}{\alpha} \hbox{ } q^{1+n})^{(n+1)^2}}{(1-\nu\hbox{ } q^n)^{(1+n)^2}(1- \nu\hbox{ } w^2\hbox{ } q^{2+n})^{(1+n)^2}}
\end{eqnarray}
Using the relations (\ref{relcon}) it is easy to understand why this procedure works. 
The equations (\ref{relcon}) show that the single trace operators of the mesonic chiral ring satisfy:
\begin{equation}
\hbox{tr}(W_{\alpha}^1 (A B )^k )= \hbox{tr}(W_{\alpha}^2 (B A )^k) \hbox{ , } \hbox{tr}(W_{\alpha}^1W^{\alpha}_1 (A B )^k )= \hbox{tr}(W_{\alpha}^2 W^{\alpha}_2 (B A )^k)
\end{equation}
which means that we must consider only the diagonal part of the spinor superfield, namely:
\begin{equation}
W_{\alpha}\equiv W_{\alpha}^1 = W_{\alpha}^2 
\end{equation}
Hence the single trace operators of the mesonic chiral ring are:
\begin{equation}\label{gentrcon}
\hbox{tr}((A B)^k) \hbox{ , } \hbox{tr}(W_{\alpha} (A B )^k )\hbox{ , } \hbox{tr}(W_{\alpha}W^{\alpha} (A B )^k) 
\end{equation}
From (\ref{gentrcon}) it is easy to understand that whenever we know the scalar mesonic 
generating function for $N=1$, the complete one is just the one dressed as in (\ref{n1conc}).\\
Let us give a look to the expansion of the generating functions for the first few values of $N$.\\
for the case $N=1$ we have:
\begin{eqnarray}
& & g_{1,0}(q,w,\alpha) = \nonumber\\
& & 1 + \Big( 4 + \frac{w}{\alpha} + w \alpha \Big) q + \Big( 9 + w^2 + 4 \frac{w}{\alpha}+ 4w\alpha \Big) q^2 + \Big( 16 + 4w^2 + 9 \frac{w}{\alpha} + 9 w \alpha \Big) q^3 + ...\nonumber\\
\end{eqnarray} 
while for the case $N=2$ we have the generating function:
\begin{eqnarray}\label{g2conic}
g_{2,0}(q,w,\alpha)=\frac{P(q,w,\alpha)}{
(1 - q)^6(1 + q)^3 \alpha^2}
\end{eqnarray}
where
\begin{eqnarray}
P(q,w,\alpha)&=& (q w + \alpha)(1 + q w \alpha)(\alpha + q ( \alpha + q ((7 + q(3 + 4 q)) \alpha + \nonumber\\
& & (1 + q (1 + q(7 + q(3 + 4 q))))w^2 \alpha + (4 + q(3 + q(7 + q + q^2))) w(1 + \alpha^2)))) \nonumber\\
\end{eqnarray}
expanding (\ref{g2conic}) we obtain:
\begin{eqnarray}
g_{2,0}(q,w,\alpha) &=& 1 + \Big( 4 + \frac{w}{\alpha} + w \alpha \Big) q + \Big( 19 + 2w^2 + 8 \frac{w}{\alpha}+ 8w\alpha \Big) q^2 + \nonumber \\ 
& & \Big( 52 + 16w^2 + 4 \frac{w^2}{\alpha^2} + 4w^2 \alpha^2 + 34 \frac{w}{\alpha} + 34 w \alpha  +\frac{w^3}{\alpha} + w^3 \alpha \Big) q^3 + ... \nonumber
\end{eqnarray}  
Before checking these expansions against the field theory operators let us make some 
comments regarding the distinction between the $U(N)$ and the $SU(N)$ gauge groups case.


\subsubsection{U(N) vs. SU(N) gauge groups}

Till now we didn't pay too much attention to the differences between the $U(N)$ groups and the $SU(N)$ groups. 
Indeed our generating functions are exact if we are going to consider field theories that have as gauge group 
a product of $SU(N)$ factors times an overall $U(1)$ factor. The presence of a $U(1)$ factor can be easily seen: 
for example the generating functions count also operators in factorized form: Tr$(W_{\alpha})$$(...)$. 
These ones would clearly be absent if the gauge group was just a product of $SU(N)$ factors. 
The additional $U(1)$ usually allows us to write down easily the generating functions. 
However if we want to count the supersymmetric degrees of freedom of a quiver gauge theory 
dual to an $AdS_5 \times H$ gravity background, we must eliminate the $U(1)$ from the generating functions, 
because no supergravity state is charged under this factor.\\
To implement the counting procedure for the $SU(N)$ groups we want essentially to impose the constraints:
\begin{eqnarray}\label{conSUN}
\hbox{ Tr }(W_{\alpha})=0 \nonumber \\
\Big( \hbox{ Tr }(W_{\alpha}W^{\alpha}) \Big)^N = 0
\end{eqnarray} 
The first constraint comes from the fact that the matrices in the algebra of $SU(N)$ groups are traceless, 
while the second one is due to a relation in the classical chiral ring. \\

\subsubsection{The $SU(N)$ conifold's mesonic chiral ring}
To implement the first constraint in (\ref{conSUN}), in the specific case of the conifold mesonic chiral ring, 
it is enough to make the substitution:
\begin{equation}\label{g1fsu}
g_{1,0}^{w,SU(N)}(q,w,\alpha) \equiv g_{1,0}^w(q,w,\alpha) - w q\Big(\alpha + \frac{1}{\alpha} \Big)
\end{equation}
while we don't have to worry about the second constraint in (\ref{conSUN}) 
because in the case of $N=1$ the $g_{1,0}$ functions count the single traces in the limit $N \rightarrow \infty$, 
and we don't need to impose any other constraints. \\
The complete $SU(N)$ generating function for $B=0$, $N=1$ is:
\begin{equation}\label{g1su}
g_{1,0}^{SU(N)}(q,w,\alpha) \equiv g_{1,0}^0(q) + g_{1,0}^{w,SU(N)}(q,w,\alpha)+ g_{1,0}^{w^2}(q,w)
\end{equation}
Consider the first few terms in its expansion:
\begin{eqnarray}\label{expg1c}
g_{1,0}^{SU(N)}(q,w,\alpha) = 1 + 4 q + \Big( 9 + w^2 + 4 \frac{w}{\alpha}+ 4w \alpha \Big) q^2 + \Big( 16 + 4w^2 + 9 \frac{w}{\alpha} + 9 w \alpha \Big) q^3 + ...\nonumber\\
\end{eqnarray} 
 The corresponding field theory gauge invariant operators are:
\begin{eqnarray}
 1 &\rightarrow& \mathbb{I} \hbox{ : } \hbox{ 1 operator } \nonumber\\ 
4 q  &\rightarrow& \hbox{Tr}(A_iB_j) \hbox{ : } \hbox{ 4 operators } \nonumber\\ 
9 q^2  &\rightarrow& \hbox{Tr}(A_iB_jA_kB_{\rho})  \hbox{ : }  \hbox{ 9 operators } \nonumber\\ 
w^2 q^2   &\rightarrow& \hbox{Tr}(W_{\alpha}W^{\alpha}) \hbox{ : }  \hbox{ 1 operator } \nonumber\\ 
\Big( 4 \frac{w}{\alpha}+ 4w\alpha \Big) q^2 &\rightarrow& \hbox{Tr}(W_{\alpha}A_iB_j)  \hbox{ : } \hbox{ 4 Weyl spinor operators } \nonumber\\
16 q^3  &\rightarrow& \hbox{Tr}(A_iB_jA_kB_{\rho}A_{\mu}B_{\nu}) \hbox{ : }  \hbox{ 16  operators } \nonumber\\ 
4w^2 q^3  &\rightarrow&  \hbox{Tr}(W_{\alpha}W^{\alpha} A_i B_j)  \hbox{ : }  \hbox{ 4  operators } \nonumber\\
\Big( 9 \frac{w}{\alpha} + 9 w \alpha \Big) q^3  &\rightarrow& \hbox{Tr}(W_{\alpha} A_i B_j A_k B_{\rho})  \hbox{ : }  \hbox{ 9 Weyl spinor operators}\nonumber\\
\end{eqnarray}
Let us pass to analyze the $N=2$ case. The first constraint in (\ref{conSUN}) is implemented 
just taking the $PE$ of the $g_{1,0}^{SU(N)}$ defined in (\ref{g1su}). We must now implement 
also the second constraint in (\ref{conSUN}). This is done applying the generalized $PE$ procedure 
explained in Section \ref{gentheo} to the $g_{1,0}^{SU(N)}$ function in (\ref{g1su}) obtained from 
(\ref{g1fsu}) and subtracting at the end the quantum numbers of operators like:
\begin{equation}
\Big( \hbox{ Tr } (W_{\alpha}W^{\alpha}) \Big) ^2 \epsilon\epsilon ((AB)^k)((AB)^j) \hbox{  }\rightarrow \hbox{  } q^4 w^4\hbox{  } \frac{1}{2}\Big(g_{1,0}^0(q^2)+(g_{1,0}^0(q))^2 \Big)
\end{equation}
as result we obtain:
\begin{equation}
g_{2,0}^{SU(N)}(q,w,\alpha) = \frac{1}{2}\partial^2_{\nu}PE[g_{1,0}^{SU(N)}(q,w,\alpha)]\Big|_{\nu = 0}- q^4 w^4\hbox{  } \frac{1}{2}\Big(g_{1,0}^0(q^2)+(g_{1,0}^0(q))^2 \Big)
\end{equation}
expanding this generating function we have:
\begin{eqnarray}
& & g_{2,0}^{SU(N)}(q,w,\alpha) = \nonumber\\
& & 1 + 4 q + \Big( 19 + w^2 + 4 \frac{w}{\alpha}+ 4w\alpha \Big) q^2 + \Big( 52 + 8w^2 + 25 \frac{w}{\alpha} + 25 w \alpha \Big) q^3 + ... \nonumber\\
\end{eqnarray}  
The first two terms are equal to the ones in the case $N=1$ in equation (\ref{expg1c}), 
then starting from $R$-charge $2$ we have more operators:
\begin{eqnarray}
19 q^2 &\rightarrow& \hbox{Tr}(A_iB_jA_kB_{\rho}) \hbox{ , } \hbox{Tr}(A_iB_j)\hbox{Tr}(A_kB_{\rho}) \hbox{ : } \nonumber\\
& & \hbox{ 9 + 10 operators} \nonumber\\
w^2 q^2 &\rightarrow& \hbox{Tr}(W_{\alpha}W^{\alpha})\hbox{ : } \nonumber\\
& & \hbox{ 1 operator} \nonumber\\
\Big( 4 \frac{w}{\alpha}+ 4w\alpha \Big) q^2 &\rightarrow& \hbox{Tr}(W_{\alpha}A_iB_j) \hbox{ : } \nonumber\\
& &\hbox{ 4 Weyl operators} \nonumber\\
52 q^3 &\rightarrow& \hbox{Tr}(A_iB_jA_kB_{\rho}A_{\mu}B_{\nu}) \hbox{ , } \hbox{Tr}(A_iB_jA_kB_{\rho})\hbox{Tr}(A_{\mu}B_{\nu})  \hbox{ : } \nonumber\\
& & \hbox{ 16 + 36 operators} \nonumber\\
8 w^2 q^3 &\rightarrow& \hbox{Tr}(W_{\alpha}W^{\alpha} A_i B_j) \hbox{ , } \hbox{Tr}(W_{\alpha}W^{\alpha})\hbox{Tr}(A_i B_j)  \hbox{ : } \nonumber\\
& & \hbox{ 4 + 4 operators} \nonumber\\
\Big( 25 \frac{w}{\alpha} + 25 w \alpha \Big) q^3  &\rightarrow&  \hbox{Tr}(W_{\alpha}A_iB_jA_kB_{\rho}) \hbox{ , } \hbox{Tr}(W_{\alpha}A_iB_j)\hbox{Tr}(A_kB_{\rho}) \hbox{ : } \nonumber\\
& & \hbox{ 9 + 16  Weyl operators} \nonumber\\
\end{eqnarray}
Now that we have understood the basic stuff of the mesonic chiral ring let us pass to 
the more interesting case of the baryonic one.

\subsection{The baryonic conifold's chiral ring}

We would like to write down the complete generating function for the chiral ring 
of the conifold theory containing all the $\frac{1}{2}$ $BPS$ degrees of freedom 
of the theory: namely the mesonic sector ($B=0$), all the operators charged under 
the $U(1)$ baryonic symmetry and of course all the possible fermionic degrees of freedom $W_{\alpha}$.

\subsubsection{The $B=1$ baryonic sector}\label{B1N2con}

Let us start analyzing the problem in the easy case of fixed baryonic charge $B$, 
and for simplicity we will analyze the sector $B=1$.\\
The prescription given in Section \ref{gentheo} basically says that we just need 
to know the $g_{1,1}^0(q)$ generating function for the scalar chiral ring. 
This one was computed in \cite{Butti:2006au} and it is:
\begin{equation}\label{consca}
g_{1,1}^0(q)=\frac{2 q^{1/2}}{(1-q)^3}
\end{equation}
Now we have to dress it with usual ``supermultiplet'' charges: 
\begin{equation}\label{g1n1b1}
g_{1,1}(q,w,\alpha)=g_{1,1}^0(q)\Big(1+q w \Big( \alpha +\frac{1}{\alpha} \Big) +  q^2 w^2 \Big)= \frac{2 \big(q^{1/2}+ q^{3/2}w(\alpha +\frac{1}{\alpha})+  q^{5/2} w^2 \big)}{(1-q)^3}
\end{equation}
The meaning of this procedure is easily explained in the case of the conifold. 
Using the relations (\ref{relcon}) we understand that the operators in the chiral 
ring in the case $N=1$ are\footnote{Observe that here we are using the real $N=1$ 
counting and hence the operators are no more matrices but numbers.}:
\begin{equation}\label{basab}
A_{i_1}B_{j_1}A_{i_2}B_{j_2}...A_{i_n}B_{j_n}A_{i_{n+1}}
\end{equation}
\begin{equation}\label{wab}
W_{\alpha}A_{i_1}B_{j_1}A_{i_2}B_{j_2}...A_{i_n}B_{j_n}A_{i_{n+1}}
\end{equation}
\begin{equation}\label{wwab}
W_{\alpha}W^{\alpha}A_{i_1}B_{j_1}A_{i_2}B_{j_2}...A_{i_n}B_{j_n}A_{i_{n+1}}
\end{equation}
The generating function (\ref{consca}) counts all the operators of the form (\ref{basab}) 
while the dressing takes into account the ones in (\ref{wab},\ref{wwab}).
The first few terms in the expansion of (\ref{g1n1b1}) are:
\begin{equation}
g_{1,1}(q,w,\alpha)= 2 q^{1/2} + \Big( 6 + 2\frac{w}{\alpha} + 2 w \alpha \Big) q^{3/2} +  \Big( 12 + 6\frac{w}{\alpha} + 6 w \alpha + 2 w^2 \Big) q^{5/2} +...
\end{equation}
We can now compare these terms with the operators in the gauge theory:
\begin{eqnarray}
 2 q^{1/2} &\rightarrow& A_i \hbox{ : } \hbox{ 2 operators} \nonumber\\
 6 q^{3/2} &\rightarrow&  A_i B_j A_k \hbox{ : } \hbox{ 6  operators} \nonumber\\
 \Big( 2\frac{w}{\alpha} + 2 w \alpha \Big) q^{3/2}  &\rightarrow& W_{\alpha}A_i  \hbox{ : } \hbox{ 2 Weyl spinor operators} \nonumber\\
 12 q^{5/2}  &\rightarrow&  A_i B_j A_k B_{\rho}A_{\mu}  \hbox{ : } \hbox{ 12 operators} \nonumber\\
\Big( 6\frac{w}{\alpha} + 6 w \alpha \Big) q^{5/2} &\rightarrow&  W_{\alpha}A_iB_j A_k  \hbox{ : } \hbox{ 6 Weyl spinor operators} \nonumber\\
2 w^2 q^{5/2} &\rightarrow&  W_{\alpha}W^{\alpha}A_i  \hbox{ : } \hbox{ 2 operators} 
\end{eqnarray}
Now we want to implement the finite $N$ counting. To reach this task we apply the rules explained in Section \ref{gentheo}.
\begin{eqnarray}
& & g_{1}(q,w,\alpha;\nu)=\sum_{\nu=0}^{\infty} \nu ^N g_{N,1}(q,w,\alpha)=\nonumber\\
& & PE[g_{1,1}(q,w,\alpha)]=PE^{\mathcal{B}}[g_{1,1}^0(q)+g_{1,1}^{w^2}(q,w)]PE^{\mathcal{F}}[g_{1,1}^w(q,w,\alpha)]\nonumber\\
\end{eqnarray}
In the case $N=2$ we have the following generating function:
\begin{eqnarray}\label{geng12yiu}
 g_{2,1}(q,w,\alpha) =
\frac{P(q,w,\alpha)}{(1 - q)^6(1 + q)^3 \alpha^2}
\end{eqnarray}
where
\begin{eqnarray}
P(q,w,\alpha)&=& q(q w + \alpha)(1 + q w \alpha)(3 \alpha + q((3 + q(9 + q)) \alpha + \nonumber\\
& & q(3 + q (3 + q(9 + q))) w^2 \alpha + (1 + 3 q(3 + q + q^2)) w (1 + \alpha^2)))\nonumber\\
\end{eqnarray}
\nonumber\\
expanding (\ref{geng12yiu}) we obtain:
\begin{eqnarray}\label{g12un}
& & g_{2,1}(q,w,\alpha) = \nonumber\\
& & 3q+ \Big( 12 + 4 \frac{w}{\alpha}+ 4 w \alpha \Big) q^2 + \Big( 45 + 8 w^2 + \frac{w^2}{\alpha^2} + 24 \frac{w}{\alpha}+ 24 w \alpha + w^2 \alpha^2 \Big) q^3 + \nonumber\\
& &  \Big( 112 + 48 w^2 + 12 \frac{w^2}{\alpha^2} + 84 \frac{w}{\alpha} + 4 \frac{w^3}{\alpha}+ 84 w \alpha + 4 w^3 \alpha + 12 w^2 \alpha^2 \Big)q^4 + ...\nonumber\\
\end{eqnarray}
and this one nicely agrees with the field theory counting:
\begin{eqnarray}\label{fieB1N2}
 3q &\rightarrow& \epsilon \epsilon (A_i)(A_j) \nonumber\\
& &\hbox{ 3 operators } \nonumber\\
12 q^2 &\rightarrow& \epsilon \epsilon (A_i B_j A_k)(A_{\rho})  \nonumber\\
& & \hbox{ 12 operators } \nonumber\\ 
\Big( 4 \frac{w}{\alpha}+ 4 w \alpha \Big) q^2  &\rightarrow& \epsilon \epsilon (W_{\alpha} A_i)(A_j)  \nonumber\\
& & \hbox{ 4 Weyl spinor operators } \nonumber\\
45 q^3 &\rightarrow& \epsilon \epsilon (A_i B_j A_k)(A_{\rho}B_{\mu}A_{\nu}) \hbox{ , } \epsilon \epsilon (A_i B_j A_kB_{\rho}A_{\mu})(A_{\nu})  \nonumber\\ & & \hbox{ 21 + 24 operators }  \nonumber\\ 
8 w^2 q^3  &\rightarrow& \epsilon \epsilon (W_{\alpha}W^{\alpha} A_i)(A_j) \hbox{ , } \epsilon \epsilon (W_{+} A_i)(W_{-}A_j)   \nonumber\\ 
& & \hbox{ 4 + 4 operators} \nonumber\\ 
 \frac{w^2}{\alpha^2}q^3  &\rightarrow& \epsilon \epsilon (W_{-} A_1)(W_{-}A_2) \nonumber\\ 
& & \hbox{ 1 operator} \nonumber\\ 
\Big(  24 \frac{w}{\alpha}+ 24 w \alpha \Big) q^3  &\rightarrow& \epsilon \epsilon (W_{\alpha} A_i B_j A_k)(A_{\rho}) \hbox{ , } \epsilon \epsilon (W_{\alpha} A_i)( A_j B_k A_{\rho})  \nonumber\\
& &  \hbox{ 12 + 12 operators} \nonumber\\ 
w^2 \alpha^2 q^3 &\rightarrow& \epsilon \epsilon (W_{+} A_1)(W_{+}A_2)  \nonumber\\ 
& & \hbox{ 1 operator} \nonumber\\  
112 q^4  &\rightarrow& \epsilon \epsilon (A_i B_j A_kB_{\rho}A_{\mu}B_{\nu}A_{\sigma})(A_{\xi}) \hbox{ , } \epsilon \epsilon (A_i B_j A_kB_{\rho}A_{\mu})(A_{\nu}B_{\sigma}A_{\xi})  \nonumber\\
& & \hbox{ 40 + 72 operators} \nonumber\\ 
48 w^2 q^4  &\rightarrow& \epsilon \epsilon (W_{\alpha}W^{\alpha}A_i B_j A_k)(A_{\rho}) \hbox{ , } \epsilon \epsilon (W_{\alpha}W^{\alpha}A_i)(A_j B_{\rho}A_k) \hbox{ , } \nonumber\\
& & \epsilon \epsilon (W_{+}A_iB_j A_k)(W_{-}A_{\rho}) \hbox{ , } \epsilon \epsilon (W_{-}A_iB_j A_k)(W_{+}A_{\rho})  \nonumber\\
& & \hbox{ 12 + 12 + 12 + 12 operators } \nonumber\\  
12 \frac{w^2}{\alpha^2}q^4 &\rightarrow& \epsilon \epsilon (W_{-}A_iB_j A_k)(W_{-}A_{\rho})  \nonumber\\
& & \hbox{ 12 operators} \nonumber\\ 
\Big( 84 \frac{w}{\alpha} + 84 w \alpha \Big)q^4   &\rightarrow&   \epsilon \epsilon (W_{\alpha}A_iB_j A_kB_{\rho}A_{\mu})(A_{\nu}) \hbox{ , } \epsilon \epsilon (W_{\alpha}A_iB_j A_k)(A_{\rho}B_{\mu}A_{\nu}) \hbox{ , } \nonumber\\
& & \epsilon \epsilon (W_{\alpha}A_i)(A_j B_k A_{\rho}B_{\mu}A_{\nu})  \nonumber\\
& & \hbox{ 24 + 36 + 24 operators} \nonumber\\
\nonumber\\
\Big(  4 \frac{w^3}{\alpha} + 4 w^3 \alpha \Big)q^4  &\rightarrow&  \epsilon \epsilon (W_{\alpha}W^{\alpha}A_i)(W_{\alpha}A_j)   \nonumber\\
& & \hbox{ 4 Weyl spinor operators} \nonumber\\
12 w^2 \alpha^2 q^4  &\rightarrow& \epsilon \epsilon (W_{+}A_iB_jA_k)(W_{+}A_{\rho})  \nonumber\\
& & \hbox{ 12 operators}
\end{eqnarray}
where with $\epsilon \epsilon (...)(...)$ we mean the obvious contractions of the epsilon indices and the operators ones. 

\subsubsection{The $SU(N)$ gauge theory}
In this Section we want to spend some words about the generating functions written in Section \ref{B1N2con}. 
Among the operators written in (\ref{fieB1N2}) there exist someone that are factorisable in product of trace and epsilon contractions.\\
Let us use as an easy example the operator 
\begin{equation}\label{infac}
\epsilon \epsilon (W_{\alpha} A_i)(A_j)
\end{equation} 
We can write other gauge invariant operators with the same quantum numbers: 
\begin{equation}\label{fact}
\hbox{ Tr } (W_{\alpha})\epsilon \epsilon (A_i)(A_j)
\end{equation}
In (\ref{infac}) there are 4 Weyl spinor operators while in (\ref{fact}) there are 3. 
This means that among the four gauge invariant operators in (\ref{infac}) there are three 
that factorize as (\ref{fact}). Indeed it is not difficult to check that:
\begin{eqnarray}\label{trW}
\epsilon \epsilon (W_{\alpha} A_1)(A_1) &\rightarrow& \hbox{ Tr } (W_{\alpha})\epsilon \epsilon (A_1)(A_1) \nonumber\\
\epsilon \epsilon (W_{\alpha} A_2)(A_2) &\rightarrow& \hbox{ Tr } (W_{\alpha})\epsilon \epsilon (A_2)(A_2) \nonumber\\
\epsilon \epsilon (W_{\alpha} A_1)(A_2) + \epsilon \epsilon (W_{\alpha} A_2)(A_1) &\rightarrow& \hbox{ Tr } (W_{\alpha})\epsilon \epsilon (A_1)(A_2)
\end{eqnarray}
 Hence the only non factorisable operator in (\ref{infac}) is the combination:
\begin{equation}
\epsilon \epsilon (W_{\alpha} A_1)(A_2) - \epsilon \epsilon (W_{\alpha} A_2)(A_1)
\end{equation}
If we really want to count operator in the $SU(N)\times SU(N)$ chiral ring and not in 
$SU(N)\times SU(N) \times U(1) $, where the last factor is the $U(1)$ overall, 
we must impose the $SU(N)$ constraint (\ref{conSUN}) and hence put equal to zero all the operators like (\ref{trW}).\\
Actually in our counting procedure we must disregard all the operators that factorize in the form:
\begin{equation}
\hbox{ Tr } (W_{\alpha}) \epsilon \epsilon (...)(...)
\end{equation}
and the ones that factorize in the form:
\begin{equation}
\Big( \hbox{ Tr } (W_{\alpha}W^{\alpha}) \Big) ^N \epsilon \epsilon (...)(...)
\end{equation}
In the case of the conifold with $N=2$ and $B=1$ the operators to disregard 
are\footnote{We write on the left hand side the type of operator and on the right 
hand side the piece of generating function to add to the $g_{2,1}(q,w,\alpha)$ defined in Section \ref{B1N2con}}:
\\
\\
\begin{eqnarray}
\Big( \hbox{ Tr }(W_{\alpha}W^{\alpha})\Big)^2 \epsilon \epsilon (A_i...)(A_j...) &\rightarrow& -q^4w^4\frac{1}{2} \Big( g_{1,1}^0(q^2)+ (g_{1,1}^0 (q))^2 \Big) \nonumber\\
\hbox{ Tr }(W_{\alpha}) \epsilon \epsilon (A_i...)(A_j...) &\rightarrow& -q w \Big( \alpha + \frac{1}{\alpha} \Big) \frac{1}{2} \Big( g_{1,1}^0(q^2)+ (g_{1,1}^0(q))^2 \Big) \nonumber\\
\hbox{ Tr }(W_{\alpha}) \epsilon \epsilon (A_i...)(W_{\alpha}W^{\alpha} A_j...) &\rightarrow& -q w \Big( \alpha + \frac{1}{\alpha} \Big) \Big( g_{1,1}^0(q) g_{1,1}^{w^2}(q,w) \Big) \nonumber\\
\hbox{ Tr }(W_{\alpha}) \epsilon \epsilon (A_i...)(W_{\alpha}A_j...) \Big|_{singlet} &\rightarrow& -q^2 w^2 \Big( g_{1,1}^0(q) \Big)^2
\end{eqnarray}
Adding these contributions to the generating function $g_{2,1}(q,w,\alpha)$ previously defined, 
we obtain the generating function $g_{2,1}^{SU(N)}(q,w,\alpha)$ that counts the gauge invariant operators in the $SU(N)$ theory. \\
Let us give a look to the first few terms in the expansion:
\begin{eqnarray}
& & g_{2,1}^{SU(N)}(q,w,\alpha) =\nonumber\\
& & 3q + \Big( 12 +
     \frac{w}{\alpha} + w \alpha \Big)q^2 + \Big( 45 + 4 w^2 + \frac{w^2}{\alpha^2} + 12\frac{w}{\alpha} + 12 w \alpha + w^2 \alpha^2 \Big)q^3 +\nonumber\\
& & \Big(112 + 24 w^2 + 12\frac{w^2}{\alpha^2} + 39\frac{w}{\alpha} + 39w \alpha  + 12w^2 \alpha^2 \Big)q^4 + ...
\end{eqnarray}
This expansion is to be compared to the one in equation (\ref{g12un}).\\
Even if it is possible to obtain the ``pure $SU(N)$'' counting, it is of course 
more natural to continue keeping the overall $U(1)$ factor, and we will indeed continue in this way in the following.

\subsubsection{The complete generating function for the conifold}

Now we would like to write down the complete generating function for the conifold 
containing all the baryonic charges and all the fermionic degrees of freedom.
The general procedure explained in Section \ref{gentheo} tells us that the complete 
generating function $g^{c.r}(q,w,\alpha,b;\nu)$ is obtained by summing over all the 
possible baryonic charges $B$ the generalized $PE$ of the $N=1$ generating functions 
with fixed baryonic charge $g_{1,B}(q,w,\alpha)$. In the conifold case there is just 
one baryonic charge $B$ running from $-\infty$ to $+\infty$ and there are no multiplicities. 
Hence equations (\ref{peppa},\ref{loca}) become: 
\begin{equation}
g^{c.r.}(q,w,\alpha,b;\nu)= \sum_{B = -\infty}^{+ \infty}b^B PE[g_{1,B}(q,w,\alpha)]
\end{equation}
The $g_{1,B}^0(q)$ generating functions for the scalar part of the chiral ring were given in \cite{Forcella:2007wk} and they are:
\begin{equation}
g_{1,B}^0(q)= \frac{(-1 + B(-1 + q)-q) q^{B/2}}{(-1 + q)^3}\nonumber\\
\end{equation}
hence
\begin{equation}
g_{1,B}(q,w,\alpha)= g_{1,B}^0(q)\Big(1 + q w \Big( \frac{1}{\alpha} + \alpha \Big) + q^2 w^2 \Big)
\end{equation}
Let us start as usual with the $N=1$ generating function.
\begin{eqnarray}\label{g1con}
g_1^{c.r.}(q,w,\alpha,b) &=& \partial _{\nu} g^{c.r.}(q,w,\alpha,b;\nu)\Big|_{\nu=0}= \nonumber\\
& & \nonumber\\
\sum_{B = -\infty}^{+ \infty}b^B g_{1,B}(q,w,\alpha) &=& \frac{1 + q w \Big( \frac{1}{\alpha} + \alpha \Big) + q^2 w^2}{\Big( 1 - \frac{q^{1/2}}{b}\Big)^2\Big( 1 - q^{1/2}b \Big)^2}
\end{eqnarray}
This result is easily explained.
In the case $N=1$ the possible generators in the chiral ring are:
\begin{eqnarray}
A_i \hbox{ , }  B_j \hbox{ , } W_{\alpha} 
\end{eqnarray}
The theory does not have superpotential and hence the scalar chiral ring is freely generated 
and one has just to impose the relations coming from the $W^i_{\alpha}$ fields.
The first terms in the expansion of (\ref{g1con}) are:
\begin{eqnarray}
g_1^{c.r.}(q,w,\alpha,b) &=& 1 + \Big( \frac{2}{b}+ 2 b \Big) q^{1/2}+ \Big( 4 + \frac{3}{b^2} + 3 b^2 + \frac{w}{\alpha} + w \alpha \Big) q + \nonumber\\
& & \Big( \frac{4}{b^3} + \frac{6}{b} + 6 b + 4 b^3 + 2\frac{w}{b \alpha} + 2\frac{ b w}{ \alpha} + 2\frac{ w \alpha }{b}+ 2 b w \alpha \Big) q^{3/2} + \nonumber\\
& & \Big( 9 + 5\frac{1}{b^4} + 8 \frac{1}{b^2} + 8 b^2 + 5 b^4 + w^2 + 4 \frac{w}{\alpha} + 3 \frac{w}{ b^2 \alpha} + 3 \frac{ b^2  w }{\alpha}+ \nonumber \\
& &  4 w \alpha + 3 \frac{w \alpha }{ b^2 }  + 3 b^2 w \alpha \Big) q^2 + ...
\end{eqnarray}
Comparing this counting with the field theory we have:
\\
\begin{eqnarray}
1 &\rightarrow& \mathbb{I} \hbox{ : } \nonumber\\
& & \hbox{ 1 operator }\nonumber\\
 \Big( \frac{2}{b}+ 2 b \Big) q^{1/2}  &\rightarrow&  B_i \hbox{ , } A_i \hbox{ : } \nonumber\\
& & \hbox{ 2 + 2 operators} \nonumber\\
4 q  &\rightarrow&  A_i B_j  \hbox{ : } \nonumber\\
& & \hbox{ 4 operators }\nonumber\\
\Big( \frac{3}{b^2} + 3 b^2 \Big) q &\rightarrow& B_i B_j \hbox{ , } A_i A_j \hbox{ : } \nonumber\\
& & \hbox{ 3 + 3 operators} \nonumber\\
\Big( \frac{w}{\alpha} + w \alpha \Big) q &\rightarrow& W_{\alpha} \hbox{ : } \nonumber\\
& & \hbox{ 1 Weyl operator}\nonumber\\
\Big( \frac{4}{b^3} + \frac{6}{b} + 6 b + 4 b^3 \Big) q^{3/2} &\rightarrow& B_i B_j B_k \hbox{ , } A_i B_j B_k \hbox{ , } \nonumber\\
& & B_i A_j A_k \hbox{ , } A_i A_j A_k :  \nonumber\\
& &  \hbox{ 4 + 6 + 6 + 4 operators}\nonumber\\ 
\nonumber\\
\Big( 2\frac{w}{b \alpha} + 2\frac{ w \alpha }{b} + 2\frac{ b w}{ \alpha}+ 2 b w \alpha \Big) q^{3/2}&\rightarrow& B_i W_{\alpha} \hbox{ , } A_i W_{\alpha} \hbox{ : } \nonumber\\
& & \hbox{ 2 + 2 Weyl operators}\nonumber\\   
9 q^2  &\rightarrow& A_i B_j A_k B_{\rho}  \hbox{ : } \nonumber\\
& & \hbox{ 9 operators }\nonumber\\ 
\Big( 5\frac{1}{b^4} + 8 \frac{1}{b^2} + 8 b^2 + 5 b^4 \Big) q^2   &\rightarrow&  B_i B_j B_k B_{\rho} \hbox{ , }  B_i B_j A_k B_{\rho} \hbox{ , } \nonumber\\
& & A_i A_j B_k A_{\rho} \hbox{ , } A_i A_j A_k A_{\rho} \nonumber\\
& & \hbox{ 5 + 8 + 8 + 5 operators }\nonumber\\ 
t^2 b^2  &\rightarrow&  W_{\alpha}W^{\alpha} \hbox{ : }\nonumber\\
& & \hbox{ 1 operator } \nonumber\\
\Big( 4 \frac{w}{\alpha} + 4 w \alpha + 3 \frac{ b^2  w }{\alpha} + 3 b^2 w \alpha + 3 \frac{w}{ b^2 \alpha} + 3 \frac{w \alpha }{ b^2 }  \Big) q^2  &\rightarrow&  W_{\alpha} A_i B_j \hbox{ , } W_{\alpha} A_i A_j \hbox{ , }  W_{\alpha} B_i B_j \nonumber\\
& & \hbox{ 4 + 3 + 3 Weyl spinor operators}\nonumber\\
\end{eqnarray}
Now we would like to compute the generating function $g_2^{c.r.}(q,w,\alpha,b)$ for the conifold with $N=2$, 
counting all the operators in the chiral ring.\\
Using the relations:
\begin{eqnarray}
g^{c.r.}(q,w,\alpha,b ; \nu) &=& \sum_{B = -\infty}^{B = + \infty} b^B PE^{\mathcal{B}} \Big[ g_{1,B}^0 (q) \Big] PE^{\mathcal{F}} \Big[ g_{1,B}^w (q,w,\alpha) \Big]PE^{\mathcal{B}} \Big[ g_{1,B}^{w^2} (q,w) \Big] \nonumber\\
&=& \sum_{\nu = 0}^{\infty} \nu^N g_N^{c.r.}(q,w,\alpha,b) 
\end{eqnarray}
One can easily obtain:
\begin{eqnarray}\label{g2t1t2}
g_2^{c.r.}(q,w,\alpha,b) &=& \sum_{B = -\infty}^{B = + \infty} b^B g_{2,B}(q,w,\alpha,b) = \nonumber\\
& &\sum_{B = -\infty}^{B = + \infty} b^B \frac{1}{2} \partial_{\nu} ^2 \Big( PE^{\mathcal{B}}[g_{1,B}^0(q)] PE^{\mathcal{F}}[ g_{1,B}^w(q,w,\alpha)] PE^{\mathcal{B}}[g_{1,B}^{w^2}(q,w)]\Big) = \nonumber \\
& &  \sum_{B = -\infty}^{B = + \infty} b^B \frac{1}{2}\Big( g_{1,B}^0(q^2) + [g_{1,B}^0(q)]^2 + g_{1,B}^{w^2}(q^2,w^2) + [g_{1,B}^{w^2}(q,w)]^2 \nonumber\\
& & 2 ( g_{1,B}^0(q) g_{1,B}^{w^2}(q,w)) - g_{1,B}^{w}(q^2,w^2,\alpha^2) + [g_{1,B}^{w}(q,w,\alpha)]^2 +\nonumber\\
& &  2 ( g_{1,B}^{0}(q) + g_{1,B}^{w^2}(q,w) ) g_{1,B}^{w}(q,w,\alpha) \Big) 
\end{eqnarray}
The easiest way to do this computation is to pass from the chemical potentials $q$ and $b$, 
counting the $R$ charge and the baryonic charge, to the chemical potentials $t_1$, $t_2$ 
counting the number of $A_i$ and $B_i$ fields, and sum over all the $SU_1(2) \times SU_2(2)$ 
symmetric representations\footnote{The theory has indeed an $SU_1(2) \times SU_2(2)$ global 
flavor symmetry under which the fields $A_i$ transform as $(2,1)$ while the fields $B_i$ transform as $(1,2)$}:
\begin{eqnarray}\label{g2t1t2br}
 g_2^{c.r.}(t_1,t_2,w,\alpha) =\frac{P(t_1,t_2,w,\alpha)}{(1 - t_1^2)^3(1 - t_1 t_2)^3(1 - t_2^2)^3 \alpha^2}
\end{eqnarray}
where
\begin{eqnarray}
P(t_1,t_2,w,\alpha)&=&((w + \alpha)(1 + w \alpha)( \alpha + t_1^3 t_2 (-3 + t_2^2)(w + t_2^2(1 + w^2)\alpha + w \alpha^2) - \nonumber\\
& & t_1^2(-1+ 3 t_2^2)(w + t_2^2 (1 + w^2)\alpha +  w \alpha^2) + w(w \alpha + t_2^2 (1 + \alpha^2)) + \nonumber\\
& & t_1 t_2 (\alpha + w^2 \alpha - (-4 + 3t_2^2) w(1 + \alpha^2)) + t_1^4 t_2^2(-3(1 + w^2)\alpha + \nonumber\\
& & t_2^2(w + 4(1 + w^2)\alpha + w \alpha^2)) + t_1^5 t_2^3(\alpha + w(w \alpha + t_2^2 (1 + \alpha^2))))) \nonumber\\
\end{eqnarray}
It is now easy to expand (\ref{g2t1t2br}) in terms of $t_1$, $t_2$ and find an expression 
to compare with the field theory result:
\begin{eqnarray} 
& & g_2^{c.r.}(t_1,t_2,w,\alpha) = \nonumber\\
& & 1 + 2 w^2 + w^4 + \frac{w}{\alpha} + \frac{w^3}{\alpha} + w \alpha + w^3 \alpha + \nonumber\\
& & t_1 t_2 \Big( 4 \frac{w^2}{\alpha^2} + 8 \frac{w}{\alpha} + 8 \frac{w^3}{\alpha}+ 4 + 16 w^2 + 4 w^4 + 8 w \alpha +  8 w^3 \alpha + 4 w^2 \alpha^2 \Big)+ \nonumber\\ & &  t_1^2 \Big( 3 + 8 w^2 + 3 w^4 +  \frac{w^2}{\alpha^2} + 4 \frac{w}{\alpha} + 4 \frac{w^3}{\alpha}+ 4w \alpha  + 4 w^3\alpha  +  w^2 \alpha ^2 \Big)+ \nonumber\\
& & t_2^2 \Big( 3 + 8 w^2 + 3 w^4 +  \frac{w^2}{\alpha^2} + 4 \frac{w}{\alpha} + 4 \frac{w^3}{\alpha}+ 4w \alpha  + 4 w^3\alpha  +  w^2 \alpha ^2 \Big)+ \nonumber\\
& & t_1^2 t_2^2 \Big( 19 + 68 w^2 + 19 w^4 +  15 \frac{w^2}{\alpha^2} + 34 \frac{w}{\alpha} + 34 \frac{w^3}{\alpha}+ 34w \alpha  + 34 w^3\alpha  +  15w^2 \alpha ^2 \Big)+ ... \nonumber\\
\nonumber\\
\end{eqnarray}
In field theory we have:\\
at level $t_1^0t_2^0$:
\begin{eqnarray}
1 &\rightarrow& \mathbb{I} \hbox{ : } \hbox{ 1 operator }\nonumber\\
2 w^2  &\rightarrow&  \hbox{Tr }( W_{\alpha}W^{\alpha} ) \hbox{ , }  \hbox{Tr }( W_{+})  \hbox{Tr } (W_{-}) \hbox{ : } \hbox{ 1 + 1 operators }\nonumber\\
w^4  &\rightarrow& (\hbox{Tr }( W_{\alpha}W^{\alpha}))^2   \hbox{ : } \hbox{ 1 operator }\nonumber\\
\Big( \frac{w}{\alpha} + w \alpha \Big)  &\rightarrow& \hbox{Tr }( W_{\alpha}) \hbox{ : } \hbox{ 1 Weyl operator }\nonumber\\
\Big( \frac{w^3}{\alpha} + w^3 \alpha \Big) &\rightarrow&  \hbox{Tr }(W_{\alpha} W^{\alpha}) \hbox{Tr }( W_{\alpha})  \hbox{ : } \hbox{ 1 Weyl operator }\nonumber
\end{eqnarray}
\\
\\
at level $t_1t_2$:
\begin{eqnarray}
t_1 t_2 \Big( 4 \frac{w^2}{\alpha^2} + 4 w^2 \alpha^2 \Big)  &\rightarrow& \hbox{Tr }( W_{-}A_i B_j) \hbox{Tr } (W_{-})\hbox{ , } \hbox{Tr }( W_{+} A_i B_j )\hbox{Tr }( W_{+})  \hbox{ : }  \nonumber\\
& & \hbox{ 4 + 4  operators }\nonumber\\
 t_1 t_2 \Big( 8 \frac{w}{\alpha}+ 8 w \alpha \Big)  &\rightarrow& \hbox{Tr }( W_{\alpha} A_i B_j ) \hbox{ , } \hbox{Tr }( W_{\alpha} )\hbox{Tr }( A_i B_j): \nonumber\\
& &  \hbox{ 4 + 4  Weyl operators }\nonumber\\
 t_1 t_2 \Big( 8 \frac{w^3}{\alpha} +  8 w^3 \alpha \Big) &\rightarrow& \hbox{Tr }( W_{\alpha} W^{\alpha} A_i B_j) \hbox{Tr } (W_{\alpha}) \hbox{ , } \hbox{Tr }( W_{\alpha} W^{\alpha})\hbox{Tr }( W_{\alpha} A_i B_j)  \hbox{ : } \nonumber\\
& & \hbox{ 4 + 4  Weyl operators }\nonumber\\
4 t_1 t_2  &\rightarrow&  \hbox{Tr }( A_i B_j) \hbox{ : } \nonumber\\
& & \hbox{ 4  operators }\nonumber\\
16  t_1 t_2  w^2 &\rightarrow& \hbox{Tr }( W_{\alpha} W^{\alpha})\hbox{Tr }( A_i B_j) \hbox{ , } \hbox{Tr }( W_{\alpha} W^{\alpha} A_i B_j)\hbox{ , } \nonumber\\ 
& & \hbox{Tr }( W_{+}A_i B_j)\hbox{Tr }( W_{-}) \hbox{ , } \hbox{Tr }( W_{-}A_i B_j)\hbox{Tr }( W_{+})  \hbox{ : }  \nonumber\\
& & \hbox{ 4 + 4 + 4 + 4 operators }\nonumber\\
4 t_1 t_2 w^4 &\rightarrow& \hbox{Tr }( W_{\alpha} W^{\alpha} A_i B_j )\hbox{Tr }( W_{\alpha} W^{\alpha}) \hbox{ : } \nonumber\\
& &  \hbox{ 4  operators }\nonumber
\end{eqnarray}
at level $t_1^2$:
\begin{eqnarray}\label{t12}
3 t_1^2  &\rightarrow&  \epsilon \epsilon (A_i) (A_j)  \hbox{ : } \nonumber\\
& & \hbox{ 3 operators }\nonumber\\
8  t_1^2  w^2  &\rightarrow&  \epsilon \epsilon (W_{\alpha} W^{\alpha} A_i) (A_j) \hbox{ , }\epsilon \epsilon (W_{+}A_i)( W_{-} A_j) \hbox{ : }\nonumber\\ 
& &  \hbox{ 4 + 4  operators }\nonumber\\
3 t_1^2 w^4 &\rightarrow& \epsilon \epsilon (W_{\alpha} W^{\alpha} A_i) (W_{\alpha} W^{\alpha}A_j) \hbox{ : } \nonumber\\
& & \hbox{ 3 operators }\nonumber\\
 t_1^2 \Big(   \frac{w^2}{\alpha^2}+ w^2 \alpha ^2 \Big)  &\rightarrow& \epsilon \epsilon (W_{-} A_1) (W_{-} A_2 )\hbox{ , }  \epsilon \epsilon (W_{+} A_1) (W_{+} A_2 )\hbox{ : }  \nonumber\\
& & \hbox{ 1 + 1 operators }\nonumber\\
 t_1^2 \Big( 4 \frac{w}{\alpha} + 4w \alpha \Big) &\rightarrow& \epsilon \epsilon (W_{\alpha} A_i) (A_j) \hbox{ : } \nonumber\\
& & \hbox{ 4 Weyl operators }\nonumber\\
 t_1^2 \Big( 4 \frac{w^3}{\alpha}+  4 w^3\alpha  \Big)   &\rightarrow& \epsilon \epsilon (W_{\alpha} W^{\alpha} A_i) (W_{\alpha} A_j) \hbox{ : }  \nonumber\\ 
& & \hbox{ 4 Weyl operators }\nonumber
\end{eqnarray}
at level $t_2^2$ we have the same set of operators (\ref{t12}) with the fields $B_i$ in the place of the $A_i$ fields.\\
At level $t_1^2 t_2^2$ instead we have:  
\begin{eqnarray}
19 t_1^2 t_2^2  &\rightarrow& \hbox{Tr }( A_i B_j A_k B_{\rho})\hbox{ , } (\hbox{Tr }( A_i B_j))^2 \hbox{ : }  \nonumber\\
& & \hbox{ 9 + 9 operators }\nonumber\\
\nonumber\\
 68 t_1^2 t_2^2 w^2  &\rightarrow&  \hbox{Tr }( W_{\alpha} W^{\alpha} A_i B_j) \hbox{Tr } (A_k B_{\rho}) \hbox{ , }\hbox{Tr }( W_{+} A_i B_j) \hbox{Tr } ( W_{-} A_k B_{\rho}) \hbox{ , } \nonumber\\
& &  \hbox{Tr }( W_{\alpha} W^{\alpha} A_i B_jA_k B_{\rho}) \hbox{ , } \hbox{Tr }( W_{\alpha} W^{\alpha})\hbox{Tr }(  A_i B_jA_k B_{\rho}) \hbox{ , } \nonumber\\ 
& & \hbox{Tr }( W_{+} ) \hbox{Tr } ( W_{-} A_i B_j A_k B_{\rho})\hbox{ , } \hbox{Tr }( W_{-} ) \hbox{Tr } ( W_{+} A_i B_j A_k B_{\rho}) \hbox{ : } \nonumber\\
& &  \hbox{ 16 + 16 + 9 + 9 + 9 + 9 operators }\nonumber\\
19  t_1^2 t_2^2 w^4 &\rightarrow&  (\hbox{Tr }(W_{\alpha} W^{\alpha} A_i B_j))^2 \hbox{ , }\hbox{Tr }( W_{\alpha} W^{\alpha}  A_i B_jA_k B_{\rho}) \hbox{Tr }( W_{\alpha} W^{\alpha} ) \hbox{ : } \nonumber\\
& &  \hbox{ 10 + 9 operators }\nonumber\\
 t_1^2 t_2^2 \Big(  15 \frac{w^2}{\alpha^2} + 15w^2 \alpha ^2 \Big)  &\rightarrow& \hbox{Tr }( W_{+} A_i B_j A_k B_{\rho}) \hbox{Tr } ( W_{+} ) \hbox{ , } \hbox{Tr }( W_{+} A_{[i} B_j ) \hbox{Tr }( W_{+} A_k B_{\rho ]}) \hbox{ , } \nonumber\\
& & \hbox{Tr }( W_{-} A_i B_j A_k B_{\rho}) \hbox{Tr } ( W_{-} ) \hbox{ , } \hbox{Tr }( W_{-} A_{[i} B_j )\hbox{Tr }( W_{-} A_k B_{\rho ]}) \hbox{ : } \nonumber\\
& & \hbox{ 9 + 6 + 9 + 6 operators }\nonumber\\
t_1^2 t_2^2 \Big(  34 \frac{w}{\alpha}+ 34w \alpha \Big) &\rightarrow& \hbox{Tr } ( W_{\alpha} A_i B_j A_k B_{\rho}) \hbox{ , } \hbox{Tr } ( W_{\alpha} A_i B_j) \hbox{Tr } (A_k B_{\rho})) \hbox{ , } \nonumber\\
& & \hbox{Tr } ( W_{\alpha})  \hbox{Tr }( A_i B_j A_k B_{\rho})\hbox{ : } \nonumber\\
& & \hbox{ 9 + 16 + 9 operators }\nonumber\\ 
 t_1^2 t_2^2 \Big(  34 \frac{w^3}{\alpha}  + 34 w^3\alpha  \Big)  &\rightarrow& \hbox{Tr } ( W_{\alpha}W^{\alpha} A_i B_j A_k B_{\rho}) \hbox{Tr }( W_{\alpha}) \hbox{ , } \hbox{Tr } ( W_{\alpha} W^{\alpha}) \hbox{Tr } ( W_{\alpha} A_i B_j A_k B_{\rho})) \hbox{ , } \nonumber\\
& &  \hbox{Tr } ( W_{\alpha}W^{\alpha}A_i B_j)  \hbox{Tr }( W_{\alpha}A_k B_{\rho})\hbox{ : } \nonumber\\
& & \hbox{ 9 + 9 + 16 operators }\nonumber 
\end{eqnarray}
It is easy to see that the generating functions count the gauge invariant operators in the right way.

\section{Conclusions}\label{conl}

In this paper we presented a systematic way to construct the complete generating functions 
for any quiver gauge theories of which we are able to write down the scalar part of the 
generating functions: namely the infinite class of gauge theories dual to toric singularities, 
to quotient singularities, to complex cones over delPezzo surfaces, and many more.\\
We solved the problem of adding the $W_{\alpha}^i$ spinorial degrees of freedom by introducing 
a kind of superfield formalism and implementing the mixed state statistic through the introduction 
of the fermionic version of the  Plethystic exponential.\\
Right now we have a good understanding of the structure of the chiral ring of a great number of quiver 
gauge theories. Possible future developments may be a systematic study of the statistical properties 
of these gauge theories, the large quantum number behavior of the various partition functions, the phase 
structure of these theories, and maybe their application to related problems such as the holographic duals 
of these thermodynamical properties. The generating function constructed in this paper contain the information 
about the density distribution of the $BPS$ degrees of freedom of the $CFT$ and hence about the entropy and more 
generically the statistical and thermodynamical properties of quiver gauge theories. For this reason they could 
be a good starting point for a microscopic understanding of the entropy of the recently constructed $AdS$ 
black holes \cite{Gutowski:2004ez,Gutowski:2004yv,Sinha:2006sh,Sinha:2007ni}.\\
The partition functions of quiver gauge theories studied in literature are mainly based on undeformed $CFT$, 
it would be interesting to study the $BPS$ spectra of $CFT$ deformed by marginal operators, or even more their 
non conformal version\footnote{See \cite{Brini:2006ej} for a review about the generic behavior of non conformal quiver gauge theories.}.\\
We plan to study these problems in future publications. 
 
\section*{Acknowledgments}
It is a great pleasure to thank Loriano Bonora, Agostino Butti, Roberto Casero, 
Amihay Hanany, Yang-Hui He, David Vegh and especially Alberto Zaffaroni for 
illumating discussions and kind encouragement.\\
I would also like to thank Constantin Bachas, Raphael Benichou, Umut Gursoy, 
Bernard Julia, Bruno Machet, Ruben Minasian, Michela Petrini, Boris Pioline, 
Giuseppe Policastro and Jan Troost for useful discussions and for their 
wonderful hospitality in Paris.\\
Even more I heartily thank Gianna, Franco, Stefano, Roberto and Nathalie 
for making me feel at home in Paris.\\
D.~F.~ is supported in part by INFN and MIUR under  
contract 2005-024045-004, by 
the European Community's Human Potential Program
MRTN-CT-2004-005104 and by the European Superstring Network MRTN-CT-2004-512194.

\bibliographystyle{JHEP}

\end{document}